\documentclass{emulateapj}

\usepackage{graphics} 
\usepackage{epsfig}
\usepackage{natbib}
\usepackage{rotating} 
\shorttitle{The Intrinsic FUV Radiation Fields of Classical T Tauri Stars}
\shortauthors{France et al.}
\begin{document}
\title{High-resolution Ultraviolet Radiation Fields of Classical T Tauri Stars\altaffilmark{*}}
\author{
Kevin France\altaffilmark{1,5}, Eric Schindhelm\altaffilmark{2}, Edwin A. Bergin\altaffilmark{3},  Evelyne Roueff\altaffilmark{4}, and Herv{\'e} Abgrall\altaffilmark{4}
}
\altaffiltext{*}{Based on observations made with the NASA/ESA $Hubble$~$Space$~$Telescope$, obtained from the data archive at the Space Telescope Science Institute. STScI is operated by the Association of Universities for Research in Astronomy, Inc. under NASA contract NAS 5-26555.}

\altaffiltext{1}{Center for Astrophysics and Space Astronomy, University of Colorado, 389 UCB, Boulder, CO 80309, USA; kevin.france@colorado.edu}
\altaffiltext{2}{Southwest Research Institute, 1050 Walnut Street, Suite 300, Boulder, CO 80302, USA}
\altaffiltext{3}{Department of Astronomy, University of Michigan, 830 Dennison Building, 500 Church Street, Ann Arbor, MI 48109, USA}
\altaffiltext{4}{LUTH and UMR 8102 du CNRS, Observatoire de Paris, Section de Meudon, Place J. Janssen, 92195 Meudon, France}
\altaffiltext{5}{NASA Nancy Grace Roman Fellow}
\begin{abstract}
The far-ultraviolet (FUV; 912~--~1700~\AA) radiation field from accreting central stars in Classical T Tauri systems influences the disk chemistry during the period of giant planet formation.  The FUV field may also 
play a critical role in determining the evolution of the inner disk ($r$~$<$~10 AU), from a gas- and dust-rich 
primordial disk to a transitional system where the optically thick warm dust distribution has been depleted.  
Previous efforts to measure the true stellar+accretion-generated FUV luminosity (both hot gas emission lines and continua) have been complicated by a combination of low-sensitivity and/or low-spectral resolution 
and did not include the contribution from the bright Ly$\alpha$ emission line.  In this work, we present a
high-resolution spectroscopic study of the FUV radiation fields of 16 T Tauri stars whose dust
disks display a range of evolutionary states.  We include reconstructed Ly$\alpha$ line profiles and remove atomic and molecular disk emission (from H$_{2}$ and CO fluorescence) to provide robust measurements of both the FUV continuum and hot gas lines (e.g., Ly$\alpha$, \ion{N}{5}, \ion{C}{4}, \ion{He}{2}) for an appreciable sample of T Tauri stars for the first time. We find that the flux of the typical Classical T Tauri Star FUV radiation field at 1 AU from the central star is $\sim$~10$^{7}$ times the average interstellar radiation field.   The Ly$\alpha$ emission line contributes an average of 88\% of the total FUV flux, with the FUV continuum accounting for an average of 8\%.   
Both the FUV continuum and Ly$\alpha$ flux are strongly correlated with \ion{C}{4} flux, suggesting that accretion processes dominate the production of both of these components.   
On average, only $\sim$~0.5\% of the total FUV flux is emitted between the Lyman limit (912~\AA) and the H$_{2}$ (0~--~0) absorption band at 1110~\AA.   The total and component-level high-resolution radiation fields are made publicly available in machine-readable format.   
\end{abstract}
\keywords{protoplanetary disks --- stars: pre-main sequence --- ultraviolet: planetary systems}
\clearpage
\section{Introduction}

Considerable observational effort has been invested to characterize the high-energy spectra of Classical T Tauri stars (CTTS), Class-II protostars with gas-rich circumstellar environments and active accretion.  Near-ultraviolet (NUV; $\lambda$~=~1700~--~3200~\AA; 7~$>$~$h\nu$~$>$~4~eV), far-ultraviolet (FUV; $\lambda$~=~912~--~1700~\AA; 13.6~$>$~$h\nu$~$>$~7~eV), extreme-ultraviolet (EUV; $\lambda$~=~120~--~912~\AA; 100~$>$~$h\nu$~$>$~13.6~eV), and X-ray ($\lambda$~$<$ 100~\AA; $h\nu$~$>$~0.1~keV) irradiances are important ingredients for a complete picture of the chemistry and evolution of protoplanetary environments around young stars during the epoch of giant planet formation, migration, and the growth of rocky planet cores~\citep{ward97,armitage02,ida04}.  FUV molecular spectra provide unique insight into the physical conditions and composition of the warm inner-disk ($r$~$<$~10 AU) surface layer~\citep{herczeg04,france11b,schindhelm12b,france12b} and molecular outflows~\citep{walter03,herczeg06,krull07}.  NUV spectra of the Balmer continuum excess provide perhaps the most direct measure of the protostellar mass accretion rate~(e.g.; Ingleby et al. 2011b, 2013 and references therein), while detailed studies of the fluxes and profiles of FUV spectral lines~\citep{giovannelli95,lamzin96,ardila02,gunther08,ardila13} and X-ray spectra~\citep{schneider08,brickhouse10} provide constraints on the magnetospheric accretion paradigm.\nocite{ingleby11b,ingleby13} Further, the intrinsic energetic radiation field, generated from a combination of accretion processes and atmospheric activity on the protostar, provide an essential input to models of the chemistry and evolution of CTTS disks.  

{\it Disk Chemistry~--~} The strength and shape of the FUV radiation field has a strong influence on the chemical abundances of the disk, both at planet-forming radii ($r$~$<$ 10 AU; Walsh et al. 2012) and at larger radii ($r$~$>$~50 AU) where the majority of the disk mass resides (see e.g., Bergin et al. 2007 and references therein).\nocite{bergin07}  The stellar FUV continuum controls the dissociation of the most abundant disk molecules (H$_{2}$ and CO; Shull \& Beckwith 1982; van Dishoeck \& Black 1988)).  The propagation of the FUV continuum is mainly regulated by dust grains~\citep{zadelhoff03}; the processes of grain-growth and settling likely allow these photons to penetrate deeper into the disk as the protoplanetary environment evolves~\citep{aikawa06,vasyunin11}.  \nocite{walsh12,shull82,vdb88}

\citet{bergin03} first emphasized the importance of accretion-generated \ion{H}{1} Ly$\alpha$ to the disk chemistry, and more recently it has been shown that the FUV spectral energy distribution of $all$ CTTSs is overwhelmingly dominated ($\gtrsim$~80\%) by Ly$\alpha$ emission~\citep{schindhelm12b}.  Unlike the FUV continuum emission, the radiative transfer of Ly$\alpha$ photons is controlled mainly by resonant scattering in the upper, atomic disk atmosphere~\citep{bethell11}.  Subsequent detailed disk modeling has demonstrated the importance of properly accounting for Ly$\alpha$ radiation from the central star, finding significant ($\gtrsim$ 1 order of magnitude) depletions in the abundances of C$_{2}$H$_{4}$, CH$_{4}$, HCN, NH$_{3}$, and SO$_{2}$ when Ly$\alpha$ is included~\citep{fogel11}.  Interestingly, some species with large photo-absorption cross-sections at Ly$\alpha$ ($\lambda$~=~1216~\AA), such as H$_{2}$O, do not show significant depletion because the enhanced dissociation rate is balanced by Ly$\alpha$-driven photodesorption of water molecules from dust grains.  It is clear now that Ly$\alpha$ is a mandatory component of FUV radiation fields used for chemical modeling.  However most large CTTS spectral atlases in the literature do not provide spectral coverage at 1216~\AA\ (e.g., Yang et al. 2012), or are dominated by geocoronal emission, such as in archival data from the {\it International Ultraviolet Explorer}.  When Ly$\alpha$ spectral coverage is included, scattering in the interstellar and circumstellar environment prevents a direct measurement of the local Ly$\alpha$ environment (as with measurements from the {\it Hubble Space Telescope}-Space Telescope Imaging Spectrograph; $HST$-STIS).\nocite{yang12}  

{\it Disk Evolution~--~}Primordial gas disks are known to dissipate on timescales of~$\leq$~10~Myr, at which point mass accretion onto the central star halts~\citep{fedele10}.  While there is a growing body of evidence that inner gas disks can survive longer than the typical 2~--~4 Myr lifetime of inner dust disks (e.g., Salyk et al. 2009; France et al. 2012b)\nocite{salyk09,france12b}, the physical process by which the inner disk is cleared is not yet established.  Various mechanisms including photoevaporation~\citep{alexander06,gorti09} and dynamical clearing by exoplanetary systems~\citep{rice03,dodson11}, possibly aided by a magnetorotational instability~\citep{chiang07}, can reproduce certain transitional disk observations.~\nocite{hernandez07,chiang07}
Photoevaporation was initially considered for EUV photons from the central star~\citep{clarke01,alexander06}, and more recent work has demonstrated that X-rays~\citep{owen10} and FUV photons~\citep{gorti09} can also play an important role in disk dispersal.  Models that simultaneously treat FUV, EUV, and X-ray irradiation from the central star have shown that the FUV illumination can control the total evaporation rate (and hence the disk lifetime) by driving the heating at intermediate ($r$~$\sim$~3~--~30 AU) and large radii ($r$~$\geq$~100 AU; Gorti \& Hollenbach 2009).  FUV radiation also controls the gas temperature at the base of the evaporative flows through the generation of photoelectrons released by FUV-illuminated dust grains.  Grain-growth and dust settling in the disk, part of the first stages of the planet formation process, enable deeper penetration of the FUV radiation.  Therefore, planet formation itself can lead to an alteration of the temperature and chemical structure of the planet-forming region.  
\citet{ingleby11} presented a comprehensive study of the evolution of FUV (excluding Ly$\alpha$) and X-ray radiation fields over the 10 Myr lifetimes of gas disks.  However, the low-resolution data used in most previous studies suffers from molecular disk contamination (line-blending with photo-excited H$_{2}$ and CO emissions) and no previous surveys of FUV radiation fields have included a proper treatment of Ly$\alpha$ or isolation of the FUV continuum.  Therefore, most data in the literature or readily available in the archive may not be representative of the strength $or$ the shape of the true disk-dispersing radiation fields. 


{\it This Work~--~} In order to provide a more accurate and complete (including local Ly$\alpha$ emission profiles) observational basis for models of disk chemistry and evolution, we present new measurements of the FUV line and continuum spectra generated by accretion and magnetic processes near the protostellar photosphere.  These FUV radiation fields are available to the community in a machine-readable format\footnote{ {\tt http://cos.colorado.edu/$\sim$kevinf/ctts\_fuvfield.html} }.   These spectra were obtained with the Cosmic Origins Spectrograph (COS) and the STIS aboard $HST$.  While {\it Far-Ultraviolet Spectroscopic Explorer} ($FUSE$) observations of a small number of bright CTTSs are available in the literature~\citep{wilkinson02,herczeg04,herczeg05,herczeg06,gunther08}, $FUSE$ did not have the sensitivity to study ``typical'' Taurus-Auriga CTTSs in detail.  We therefore use $FUSE$ observations to constrain our 912~--~1150~\AA\ radiation field creation (in particular the shapes and strengths hot gas emission lines and continua), but we present an approach where the short-wavelength radiation fields are inferred from longer-wavelength $HST$ data.  

\begin{figure}
\figurenum{1}
\begin{center}
\epsfig{figure=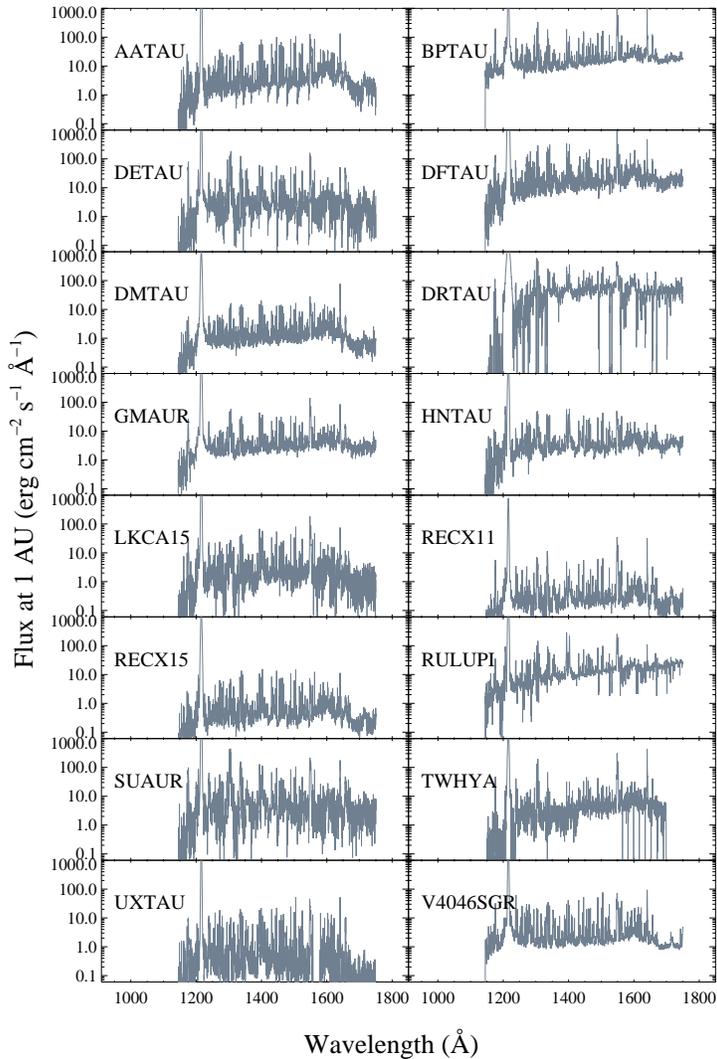,width=3.4in,angle=00}
\vspace{+0.25in}
\caption{Complete FUV spectra of the 16 CTTSs studied in this work, including reconstructed Ly$\alpha$ emission lines.
These spectra are coadditions of $HST$-COS observations in the G130M and G160M modes (except for TW Hya, which was observed with STIS E140M; Herczeg et al. 2002) at several central wavelengths and focal-plane split positions.  Almost all of the structures seen in these data are real atomic and molecular emission and absorption features.
The data have been corrected for interstellar reddening (Table 1), scaled to the flux at 1 AU from the central star for comparison, and binned by three spectral resolution elements (21 pixels) for display.  
\label{cosovly}}
\end{center}
\end{figure}

\begin{figure}
\figurenum{2}
\begin{center}
\epsfig{figure=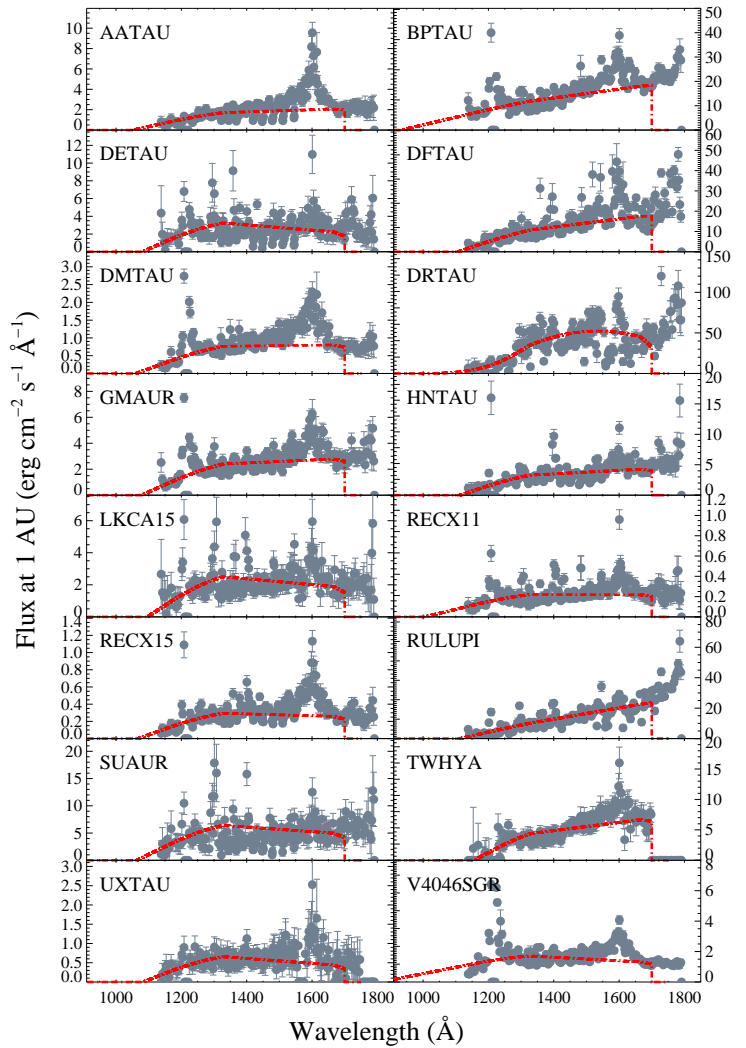,width=3.4in,angle=00}
\vspace{+0.25in}
\caption{The binned FUV continuum spectra are shown as gray filled circles.  A second order polynomial fit is extrapolated down to the Lyman Limit (912~\AA) and is shown as the red dashed line.   The ``1600~\AA\ Bump'' (spanning $\sim$~1520~--~1660~\AA) is prominent (detected at $>$ 3$\sigma$ significance) in  10/16 targets.     
\label{cosovly}}
\end{center}
\end{figure}

In Section 2, we describe the targets and the $HST$ observations.  In Section 3, we briefly describe the spectral deconvolution performed to separate the continuum, hot gas atomic line, and molecular line emission.   The details of this deconvolution can be found in the Appendix.  We detect the quasi-continuous emission feature near 1600~\AA, the ``1600~\AA\ Bump'' in $\sim$~70\% of the sources; the analysis of this emission~\citep{bergin04,ingleby09,france11a}  will be the subject of a future work.  In Section 4, we discuss the stellar+accretion spectra and present correlations suggesting that the majority of the Ly$\alpha$ and FUV continuum are generated by accretion processes.  
We present a summary of this work in Section 5.


\section{$HST$ Targets and Observations} 

We present 16 objects from the larger GTO + DAO T Tauri star samples described by Ardila et al. (2013; focusing on the hot gas emission lines) and France et al. (2012; focusing on the molecular circumstellar environment).  The observations presented here have been described in detail in these previous works, and we briefly summarize the data.  
11 of the 16 sources were observed as part of the DAO of Tau guest observing program (PID 11616; PI - G. Herczeg), 4 were part of the COS Guaranteed Time Observing program on protoplanetary disks (PIDs 11533 and 12036; PI - J. Green), and we have included archival STIS observations of the well-studied CTTS TW Hya~\citep{herczeg02,herczeg04}, obtained through StarCAT~\citep{ayres10}.  The targets were selected by the availability of reconstructed Ly$\alpha$ spectra, as this 
emission line is a critical component to the intrinsic CTTS UV radiation field~\citep{schindhelm12b} and has not been uniformly included in recent studies of the CTTS radiation field (e.g., Ingleby et al. 2011; Yang et al. 2012).\nocite{ingleby11,yang12}  

\begin{figure}
\figurenum{3}
\begin{center}
\epsfig{figure=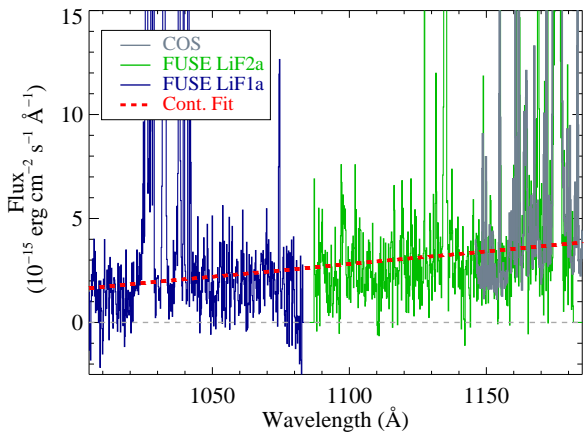,width=2.5in,angle=90}
\vspace{-0.1in}
\caption{A comparison between the extrapolation of the V4046 Sgr FUV continuum fit into the $FUSE$ spectral range (1005~--~1185~\AA).  The gray solid line is the $HST$-COS data and the red dashed line is the continuum fit (see Section 3 for an overview and Appendix A.3 for details).  The green solid line shows a coaddition of $FUSE$ LiF 2a channel data and the blue solid lines shows a coaddition of $FUSE$ LiF 1a channel data.  The $HST$-based extrapolation is a satisfactory representation of the short-wavelength FUV continuum flux observed in the archival $FUSE$ spectra.  
\label{cosovly}}
\end{center}
\end{figure}

\begin{deluxetable*}{lccccc}
\tabletypesize{\normalsize}
\tablecaption{$HST$ Target List \label{lya_targets}}
\tablewidth{0pt}
\tablehead{ \colhead{Target\tablenotemark{a}}   &   \colhead{A$_V$}    &   \colhead{Distance}    &     \colhead{$\dot M$}      &   \colhead{$HST$ ID\tablenotemark{a}}   &   \colhead{Ref.\tablenotemark{b}}  \\ 
   &   &    (pc)    &           (10$^{-8}$ M$_{\odot}$ yr$^{-1}$)        &   (ks)   &    } 
\startdata
AA Tau    &    0.5   &  140    & 0.33  &    11616    & 2,8,17   \\  
BP Tau    &    0.5  &  140    &  2.88  &  12036       & 1,8      \\  
DE Tau    &    0.6  &   140    &  2.64  &    11616   & 1,8       \\  
DF Tau    &    0.6   &   140     &  17.7  &   11533   & 1,8,14      \\  
DM Tau    &    0.0   &  140     &  0.29  &   11616   & 2,17,18      \\  
DR Tau    &    1.2  &  140     &  3.16  &    11616   & 2,10,15      \\  
GM Aur    &    0.1  &   140     &  0.96  &  11616     & 2,8,17        \\  

HN Tau    &    0.5   &  140    &  0.13      & 11616  &  1,8       \\  

LkCa 15    &    0.6  &  140    &  0.13  &    11616   &  1,9    \\  
RECX 11    &    0.0   &   97     &  0.03  &    11616 &  3,16     \\  
RECX 15    &    0.0   &   97    &  0.10  &    11616 &  4,12,16   \\  
RU Lupi    &    0.07   &  121    &  3.00  &    12036 & 5,19   \\  

SU Aur    &    0.9   &   140     &  1.0  &    11616    &  1,10        \\ 
TW Hya    &    0.0   &   54     &  0.02  &    8041$-S$  &     6,11 \\ 

UX Tau    &    0.20   &   140     &  1.00  &    11616  &  1,13   \\  
V4046 Sgr    &    0.0  &   83    &  1.30  &   11533   & 7,14 
\enddata
\tablenotetext{a}{{\it - S} denotes that the observation was made with $HST$-STIS.  All other observations were obtained with $HST$-COS.  }
\tablenotetext{b}{\rule{0mm}{5mm}
(1) \citet{Kraus2009}; (2) \citet{Ricci2010}, age uncertainties are assumed to be $\pm$~0.20;   (3) \citet{Lawson2001}; (4) \citet{RHowat2007}; (5) \citet{Herczeg2005}; (6) \citet{Webb1999}; (7) \citet{Quast2000}; (8) \citet{Gullbring1998}; (9) \citet{Hartmann1998}; (10) \citet{Gullbring2000}; (11) \citet{Herczeg2008};  (12) \citet{Lawson2004}; (13) \citet{Andrews2011}; (14) \citet{france11a}; (15) \citet{White2001};  (16) \citet{ingleby11b}, 
(17) \citet{kenyon95}; (18) \citet{valenti93}; (19) \citet{herczeg06}}
\end{deluxetable*}

Most of the targets were observed with the medium-resolution FUV modes of COS (G130M and G160M; Green et al. 2012).~\nocite{green12} These observations were acquired between 2009 December and 2011 September. Multiple central wavelength settings at several focal-plane split positions were used to create continuous FUV spectra from $\approx$ 1150~--~1750~\AA\ and mitigate the effects of fixed pattern noise.  These modes provide a point-source resolution of $\Delta$$v$~$\approx$~17 km s$^{-1}$ with 7 pixels per resolution element~\citep{osterman11}. 
The STIS observations were acquired using the E140M echelle mode ($\Delta$$v$~$\approx$~8 km s$^{-1}$) through the 0.2''~$\times$~0.2'' aperture.   For 6/16 targets (DM Tau, DR Tau, GM Aur, HN Tau, RECX-11, and RECX-15), we also acquired nearly simultaneous observations (made on subsequent $HST$ orbits) with the STIS G230L mode in order to measure the NUV spectrum and constrain the mass accretion rate by modeling the Balmer continuum excess~\citep{ingleby13}.  Table 1 lists the targets studied in this paper and the relevant system parameters.  

The targets are found in several nearby star-forming regions and include primordial (e.g., AA Tau, BP Tau, DF Tau), `pre-transitional' (sub-AU inner disk dust cavity; V4046 Sgr), and transitional CTTS systems (e.g., GM Aur, DM Tau, LkCa15, UX Tau A).  In order to compare the strength of the radiation fields in the inner disk in this paper, we present plots of the various emissions scaled to 1 AU from the central star.  The observed spectra, corrected for interstellar reddening, are shown in Figure 1.   We assumed the following distances for our sources: For the Taurus-Auriga targets (AA Tau, BP Tau, DE Tau, DF Tau, DM Tau, DR Tau, GM Aur, HN Tau, LkCa15,  SU Aur, UX Tau A), we assumed $d$~=~140 pc~(Elias 1978; Kenyon \& Hartmann 1995; and see also the VLBA work presented by Loinard et al. 2007); for the $\eta$ Cha targets (RECX-11, RECX-15), we assumed $d$~=~97 pc~\citep{mamajek99}; and for the TW Hya association, we assumed $d$~=~54 pc~\citep{leeuwen07}.  Other objects are V4046 Sgr ($d$~=~83 pc; Quast et al. 2000) and RU Lupi ($d$~=~121 pc; van Leeuwen et al. 2007).  
\nocite{quast00,comeron03,feigelson03,leeuwen07,herczeg05,elias78,kenyon95,loinard07,kraus12}


\begin{figure}
\figurenum{4}
\begin{center}
\epsfig{figure=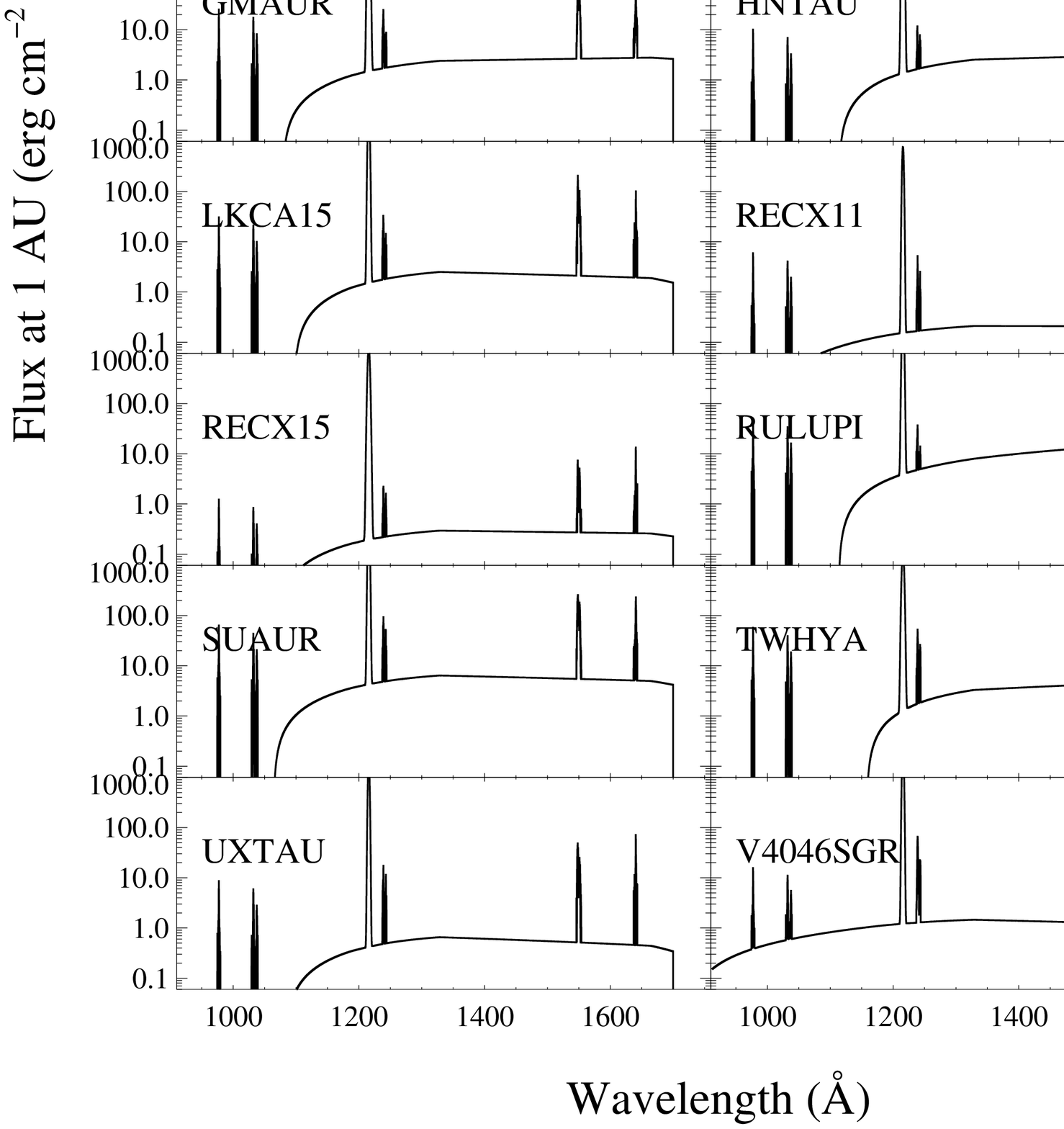,width=3.4in,angle=00}
\vspace{+0.25in}
\caption{Combined intrinsic CTTS FUV radiation fields, normalized to the flux at 1 AU from the central star for comparison.   These spectra include a polynomial fit to the observed FUV continuum, reconstructed Ly$\alpha$ radiation fields based on fluorescent H$_{2}$ emission lines, and observationally determined hot gas emission lines.  These spectra are available in machine readable format at:  {\tt  http://cos.colorado.edu/$\sim$kevinf/ctts\_fuvfield.html}.  
\label{cosovly}}
\end{center}
\end{figure}

\section{Far-Ultraviolet Radiation Fields}

When observed at moderate-to-high spectral resolution, CTTSs display some of the richest FUV emission line spectra of any astrophysical object.  The spectra include emission from several different processes, ranging from the protostellar atmosphere~\citep{gunther08,ingleby11}, accretion flows~\citep{castro96,yang11,ardila13}, molecular and atomic outflows~\citep{herczeg05,herczeg06,krull07}, as well as molecular diagnostics of the surrounding planet-forming disk~\citep{herczeg04,france11b,schindhelm12a,france12b}.  While this richness means that the inner disk+accretion system can be studied together using narrow-band photometry ($R$~$<$~100) or low-resolution spectroscopy ($R$~$\sim$~1000), the study of any individual component necessarily requires some measure of spectral decomposition in order to extract the emission (or absorption) signature of interest.  Below, we briefly describe the analysis used to extract the most important components of the intrinsic FUV radiation fields of the 16 objects in this sample.  We refer the reader to the Appendix for details of the construction of each FUV radiation field component.  


Accretion shocks are a significant source of hot gas in protostellar systems, observed as UV and X-ray emission lines in excess of what can be attributed to magnetospheric activity alone~\citep{simon90,calvet98,krull00,gunther08}. Specifically, excess emission from neutral hydrogen (line formation temperature $T_{form}$~$\sim$~10$^{4}$ K; observed as broad Ly$\alpha$ and H$\alpha$ emission) and the C$^{3+}$ ion ($T_{form}$~$\sim$~10$^{5}$ K, assuming a collisionally ionized environment\footnote{We note that \ion{C}{4} may also be produced by photoionization at lower temperatures~\citep{lamzin03,gunther08}.}; observed through the $\lambda\lambda$~1548, 1550~\AA\ \ion{C}{4} resonance doublet) correlate well with the mass accretion rate 
~\citep{white03,krull00,ardila13}. Several authors have used high-resolution spectra of the brightest CTTS, TW Hya, to separate the FUV continuum, Ly$\alpha$ emission, and hot gas lines from the molecular emission lines~\citep{costa00,bergin03,herczeg02,herczeg04}, however detailed studies of fainter CTTSs were challenging with the available suite of ultraviolet spectrographs prior to 2009 ($HST$ Servicing Mission 4).   Using the high spectral resolution and high sensitivity of COS, we have carried out this spectral decomposition for the other 15 targets presented herein.  

The 912~--~1700~\AA\ FUV radiation fields produced in this work have three components: 1) a spectrally resolved, reconstructed  Ly$\alpha$ emission profile based on the observed H$_{2}$ fluorescence spectra of our targets, 2) the H$_{2}$-subtracted, spectrally resolved hot gas emission line profiles of the relevant ionization states of helium, carbon, nitrogen, and oxygen, and 3) the disk-subtracted FUV continuum emission that is created from a polynomial fit to spectral regions free of atomic, H$_{2}$, and CO emission features (Figure 2).  Details of the Ly$\alpha$ reconstruction~\citep{schindhelm12b}, the removal of H$_{2}$ emission from hot gas lines, and the measurement of the FUV continuum (see also France et al. 2011a,b) are described in detail in the Appendix.\nocite{france11a,france11b} It is important to re-emphasize here that our primary data only extend to $\lambda$~$\approx$~1140~\AA; the radiation fields at shorter wavelengths are estimates based on existing $FUSE$ observations of CTTSs and an extrapolation of the 1140~--~1700~\AA\ continuum shape to the Lyman limit.  As described in Section A.2, we scale the hot gas lines in the $FUSE$ range (\ion{C}{3} $\lambda$977 and \ion{O}{6} $\lambda$$\lambda$1032,1038) to the hot gas lines observed in the $HST$ bandpass.  A significant uncertainty in the short-wavelength continuum is the fidelity of the extrapolation to 912~\AA.  In Figure 3, we display the spectrum of V4046 Sgr,  the source with the highest S/N $HST$ spectra from 1140~--~1275~\AA\ and with a detection of the FUV continuum from $FUSE$.  This demonstrates that the FUV continuum extrapolation to shorter wavelengths provides a reasonably good representation of the V4046 Sgr $FUSE$ spectrum, justifying our approach for the remainder of the targets.  The complete radiation fields are shown in Figure 4.  Plots showing the reconstructed Ly$\alpha$ profiles, the H$_{2}$ ``cleaning'' process for extracting hot gas line profiles, the final hot gas line profiles, 
and the extracted FUV continua are displayed in Figures A.1~--~A.6.


\begin{figure*}
\figurenum{5}
\begin{center}
\epsfig{figure=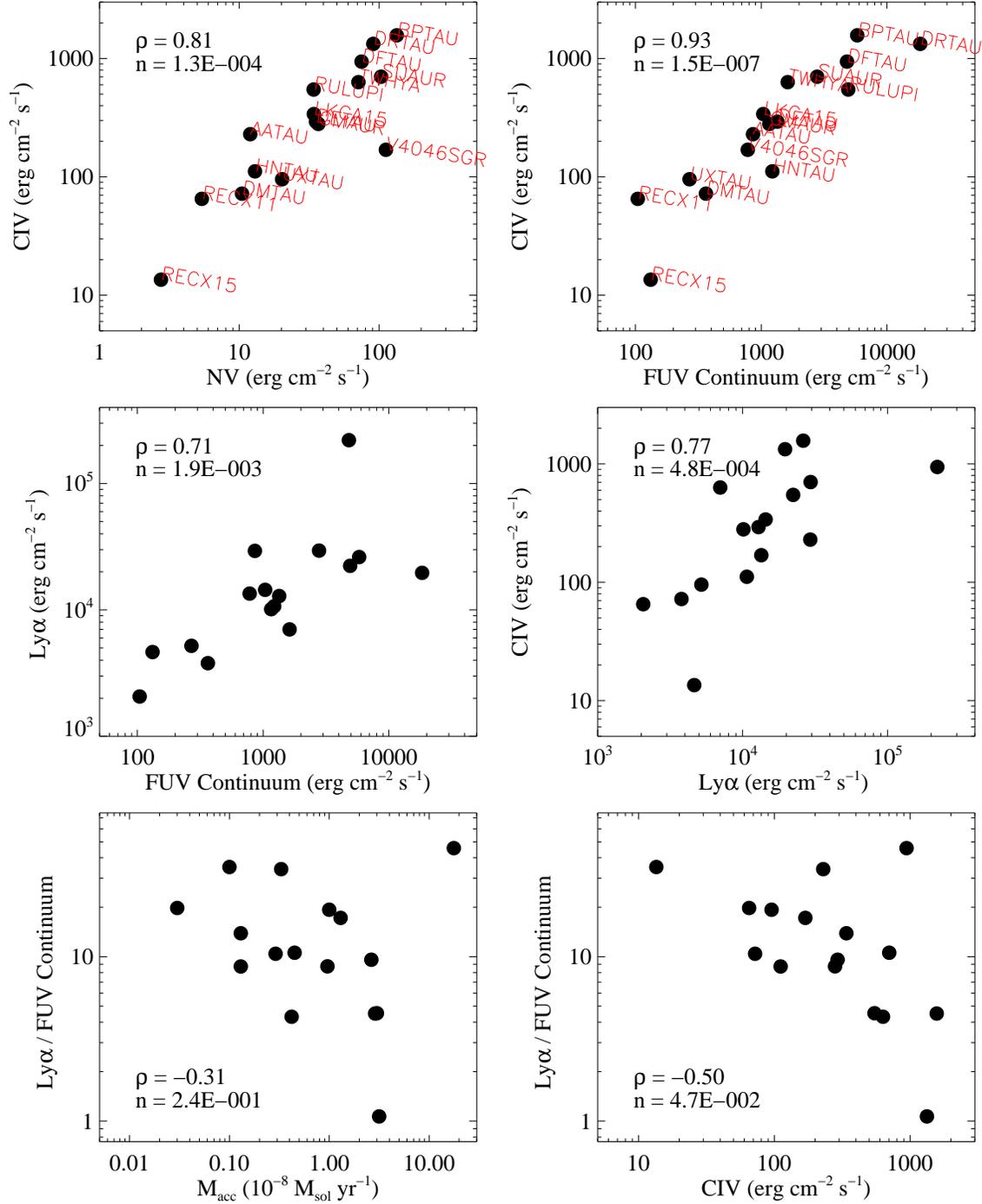,width=6.0in,angle=00}
\vspace{+0.2in}
\caption{ Correlations between emission line and continuum flux ratios.  See \S 4.1 for a discussion. The reddening-corrected fluxes are evaluated at 1 AU from their central stars for comparison.  The statistics for the significance of the fit are shown for each plot; the `$\rho$' is Spearman's rank correlation and `$n$' is a measure of the likelihood that the data are drawn from a random sample.  $\vert$$\rho$$\vert$~$>$~0.5 and $n$~$<$~0.05 indicates a likely correlation.  
\label{cosovly}}
\end{center}
\end{figure*}


\section{Discussion}

There are numerous indicators of the gas and dust content of a young protoplanetary system. Three important observables are the warm dust content of the inner disk, the presence of circumstellar gas, and signs of active accretion. Our $HST$ observations provide measurements of the last two, while the first has been extensively studied in the IR. 
The large transition probabilities of the H$_{2}$ electronic band systems and the lack of photospheric emission at $\lambda$~$<$~1700~\AA\ in low-mass stars make fluorescent H$_{2}$ one of the most sensitive indicators for the presence of molecular gas in the inner ~$\sim$~10~AU of young circumstellar disks~\citep{france12b}. H$_{2}$ emission line spectroscopy can directly probe gas surface densities as small as $\Sigma_{H2}$~$\lesssim$~10$^{-6}$ g cm$^{-2}$.  
The FUV bandpass also contains resonance lines of hydrogen, carbon, nitrogen, and oxygen which are powerful diagnostics of warm/hot gas ($T_{form}$~$\sim$~10$^{4}$~--~3~$\times$~10$^{5}$ K; \ion{H}{1} Ly$\alpha$, \ion{C}{3} $\lambda$977, \ion{C}{4} $\lambda$$\lambda$1548,1550, \ion{N}{5} $\lambda$$\lambda$1239,1243, and \ion{O}{6} $\lambda$$\lambda$1032,1038; as well as the H$\alpha$ line of ionized helium, \ion{He}{2} $\lambda$1640) formed near the accretion shock region\footnote{including the heated photosphere at the base of the accretion column, the pre-shock region, shock surface, and post-shock region}  by collisional- and photo- excitation and ionization processes.  
Our observations enable the first measurements of the FUV continuum in many of these targets, a potentially useful diagnostic of the accretion environment that has only recently been made accessible by the high sensitivity and spectral resolution of $HST$-COS (e.g., France et al. 2011a).\nocite{france11b}

\subsection{Ly$\alpha$, Hot Gas Lines, and the FUV Continuum: Constraints on Physical Origins 
}

By bringing together the reconstructed Ly$\alpha$ emission profiles, the ``H$_{2}$ cleaned'' hot gas emission lines, and measurements of the uncontaminated FUV continuum, we can explore the relations between these different components and empirically constrain the origin of the strong Ly$\alpha$ and FUV continuum emission in these sources.  The prevailing picture for the formation of the UV continuum excess is that it arises from ionized gas ($T_{cont}$~$\sim$~1.5~--~3~$\times$~10$^{4}$ K) in an optically thin pre-shock region at the base of the accretion column~\citep{calvet98,costa00}.  In this picture, the FUV continuum may be an extension of the Balmer continuum used for accretion rate determination in the NUV (see e.g., Ingleby et al. 2013); and the connection between the NUV and FUV continua has been explored previously for DF Tau~\citep{france11a}.   In the following subsection, we show that for a subsample of six of our targets with nearly simultaneous FUV and NUV observations, the FUV and NUV continua are characterized by a significantly different slope.  This demonstrates that the connection proposed by~\citet{france11a} was most likely an artifact of comparing non-simultaneous observations, although this cannot be conclusively demonstrated for all of our targets in our sample without contemporaneous FUV and NUV data.  

We find that while some of our sources display a monotonically decreasing FUV continuum towards shorter wavelengths (BP Tau, DF Tau, RU Lup, see also Herczeg et al. 2005), many of the FUV continua are flat across the FUV band (GM Aur, RECX-15, UX Tau, V4046 Sgr).  This suggests that multiple physical processes likely contribute in the complex emitting region near the stellar photosphere and the inner accretion disk.  Our aim here is to provide the data for use in detailed models of disk chemistry and evolution, and therefore we do not attempt to present a comprehensive model for the continuum emission region.  

The hot gas emissions from \ion{C}{4} and \ion{N}{5} have been long considered to be related to accretion processes~\citep{krull00}. A detailed spectral line analysis of a larger sample of COS and STIS observations has shown that these lines are likely produced by hot gas in accretion spots near the stellar photosphere, the edges of accretion columns, or multiple accretion columns of varying densities~\citep{ardila13}.  The \ion{He}{2} lines however appear to be dominated by a combination of emission from the magnetically active stellar atmosphere and the pre-shock region.  Ly$\alpha$ is presumably generated at several places in the near-star environment, however we are not aware of any previous work that explores the observational connection between Ly$\alpha$ luminosity and accretion processes (see Muzerolle et al. 2001 and Kurosawa et al. 2006 for a detailed description of Balmer lines).\nocite{muzerolle01,kurosawa06}
\begin{figure*}
\figurenum{6}
\begin{center}
\epsfig{figure=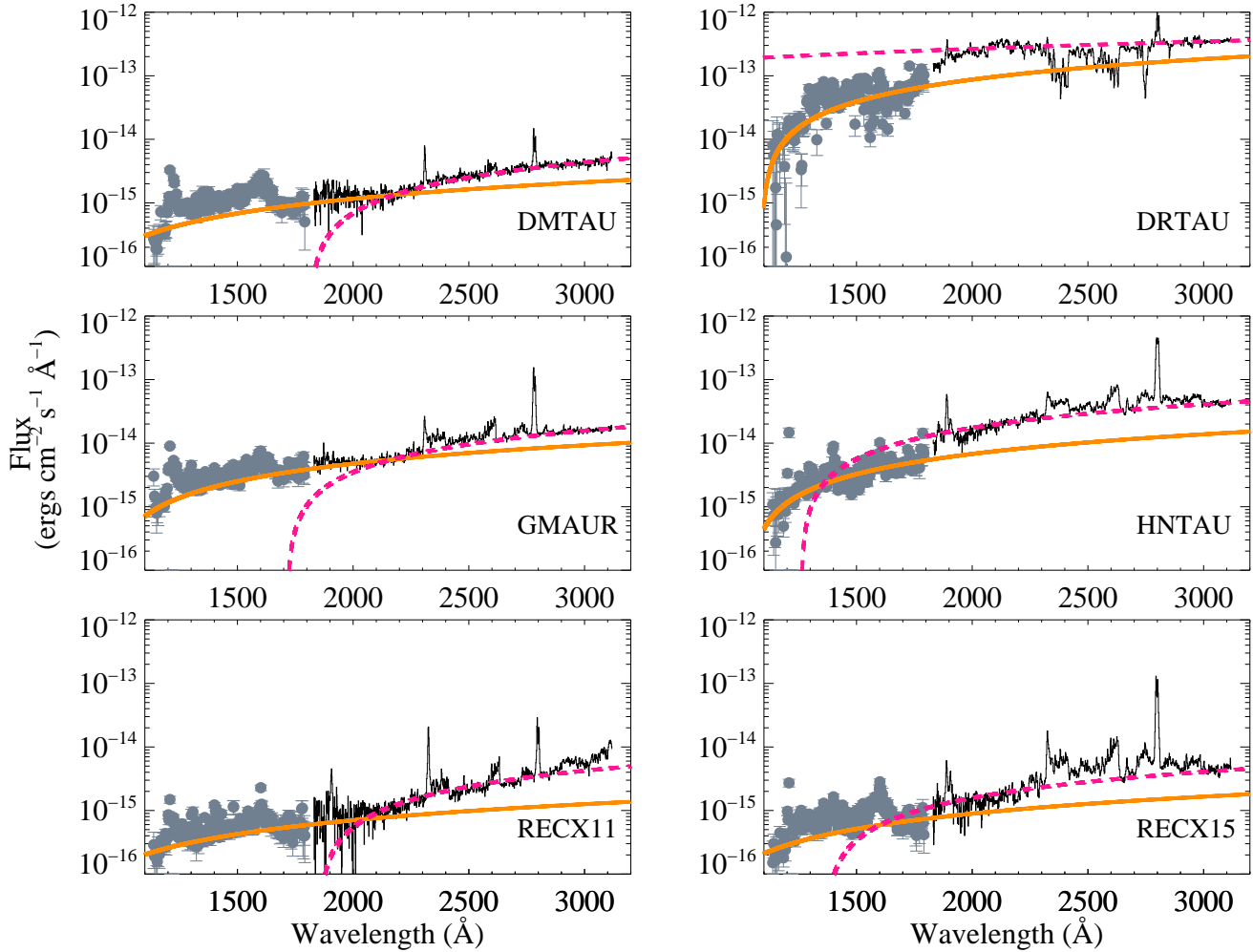,width=5.0in,angle=90}
\vspace{+0.25in}
\caption{ A comparison between the reddening corrected FUV (gray filled circles) and NUV (black line) continua for the six stars in our sample with contemporaneous observations.  The solid orange line and dashed pink lines are linear fits to the FUV and NUV continua, respectively.  The linear fit to the FUV data is created using fluxes from 1150~--~1174~\AA\ and 1680~--~1750~\AA, excluding contributions from the molecular continuum and the ``1600~\AA\ Bump''.  The linear fits show that the FUV continuum is brighter than suggested by an extrapolation of the NUV continuum for most targets, and that the FUV slopes are shallower, possibly suggesting multiple emitting regions.  
\label{cosovly}}
\end{center}
\end{figure*}

In order to empirically constrain the relationship between Ly$\alpha$ and the FUV continuum and accretion processes in the following subsections, we assume that the integrated \ion{C}{4} flux is a proxy for the mass accretion rate~\citep{krull00,ardila13}.  It is critical to have this quasi-simultaneous accretion diagnostic because accretion rates and UV fluxes from CTTSs are known to vary by factors of $\sim$~2 on timescales of months-to-years~\citep{valenti00}.  Therefore, accretion rates in the literature will not necessarily be well correlated with the UV emission at the time of our observations.  In Figure 5 {\it top left}, we demonstrate the tight correlation between the accretion-dominated \ion{C}{4} and \ion{N}{5} flux for our stars (see also the larger sample from Ardila et al. 2013).  For all of the correlations shown in Figure 5, we present the Spearman rank correlation coefficient ($\rho$) and the likelihood that this population is drawn from a random sample (i.e., not correlated; $n$).  In general,  $\vert$$\rho$$\vert$~$>$~0.5 and $n$~$<$~0.05 indicates that a real correlation exists between the two quantities.   For instance, [$\rho$,$n$] = [0.81, 1.3~$\times$~10$^{-4}$] for the plot of \ion{N}{5} and \ion{C}{4}, indicating a strong correlation, as expected.

\subsubsection{FUV Continuum}  The FUV continuum flux (in units of erg cm$^{-2}$ s$^{-1}$), evaluated at 1 AU from the central star, is computed by integrating the polynomial fit to the binned intra-emission-line $HST$ spectra (\S A.3) from 912~--~1650~\AA\ (Figures 2 and A.4~--~A.6).  The FUV continuum is shown to be tightly correlated with the \ion{C}{4} emission (Figure 5, {\it top right}, [$\rho$,$n$] = [0.93, 1.5~$\times$~10$^{-7}$]).  We interpret this as strong evidence that the CTTS FUV continuum is generated by accretion processes, either directly or powered by accretion luminosity.  However, it may originate in a spatially separate region from the NUV Balmer continuum, as mentioned above.   

In Figure 6, we compare data for six of our targets (DM Tau, DR Tau, GM Aur, HN Tau, RECX-11, and RECX-15) with FUV and NUV observations separated by less than a few hours.  Linear fits to the FUV (excluding the molecular quasi-continuum) and NUV regimes are shown overplotted as solid orange and dashed pink lines, respectively.  The observed FUV continuum is shown to be brighter than predicted from a simple extrapolation of the NUV continuum, and has a significantly shallower slope (see also Herczeg et al. 2004, 2005).\nocite{herczeg04,herczeg05} 
 We note that the S/N ratios of individual binned data points 
are relatively high, demonstrating that the excess FUV continuum flux is statistically significant in all cases. The average S/N per binned continuum point ranges from 2.5~--~14.8 from 1140~--~1340~\AA\ and 0.9~--~17.0 from 1660~--~1740~\AA\ (Table A.1).  
While it seems clear that the FUV continuum is related to accretion processes, these differences argue that the FUV and NUV continua may be formed in spatially distinct regions near the interaction of the accretion streams and the stellar photosphere.    Recent models of the NUV Balmer excess have incorporated multiple accretion components with a range of energy fluxes and densities (e.g., Ingleby et al. 2013), and future work may be able to connect the FUV and NUV continua by incorporating these different components.  

Our analysis finds that the FUV continuum extends to $\lambda$~$<$~1216~\AA\ in all cases, ruling out the possibility that this flux originates from \ion{H}{1} 2-photon (2$s$~$^{2}S_{1/2}$~--~1$s$~$^{2}S_{1/2}$) emission.  Furthermore, \ion{H}{1} 2-photon emission is suppressed in high-density regions ($n_{H}$~$>$~1.5~$\times$~10$^{4}$ cm$^{-3}$), consistent with the scenario in which this emission is the high-frequency extension of the Balmer continuum generated in the relatively dense pre-shock region at the base of the accretion column~\citep{calvet98}.  A future space-borne instrument capable of simultaneous FUV and NUV spectroscopy or photometry could provide more insight on the potential link between the FUV continuum and the NUV Balmer continuum in CTTSs.

\begin{deluxetable*}{lccccccc}
\tabletypesize{\scriptsize}
\tablecaption{Relative Contributions to the Stellar + Accretion FUV Radiation Field\label{lya_lines}}
\tablewidth{0pt}
\tablehead{
\colhead{Target} & \colhead{$F_{tot}$\tablenotemark{a}} &   \colhead{log$_{10}$($F_{tot}$/G$_{o}$)\tablenotemark{b} }   & 
\colhead{FUV Continuum} & \colhead{Ly$\alpha$ \tablenotemark{c}} & \colhead{\ion{C}{4}} & \colhead{Other lines\tablenotemark{d}}  & \colhead{$\lambda$~$\leq$~1110~\AA}  \\ 
   &    (erg cm$^{-2}$ s$^{-1}$)    &      &    (\%)    &     (\%)     &    (\%)     &      (\% )  &      (\% )   }
\startdata
AATAU & 3.1~$\times$~10$^{4}$   &   7.3   &   2.8   &   95.9   &   0.8   &   0.6  &  0.1   \\  
BPTAU & 3.5~$\times$~10$^{4}$   &   7.3   &   16.8   &   75.7   &   4.5   &   3.1 &  1.8  \\  
DETAU & 1.5~$\times$~10$^{4}$   &   7.0   &   9.2   &   88.0   &   2.0   &   0.9  &  0.3 \\  
DFTAU & 2.3~$\times$~10$^{5}$   &   8.2   &   2.1   &   97.2   &   0.4   &   0.2   &  0.1\\  
DMTAU & 4.3~$\times$~10$^{3}$   &   6.4   &   8.4   &   88.0   &   1.7   &   1.9 &  0.3  \\  
DRTAU & 4.0~$\times$~10$^{4}$   &   7.4   &   46.2   &   49.4   &   3.3   &   1.0  &  0.7 \\  
GMAUR & 1.2~$\times$~10$^{4}$   &   6.9   &   9.9   &   86.2   &   2.4   &   1.5  &  0.4 \\  
HNTAU & 1.2~$\times$~10$^{4}$   &   6.9   &   10.1   &   88.4   &   0.9   &   0.5 &  0.1  \\  
LKCA15 & 1.6~$\times$~10$^{4}$   &   7.0   &   6.5   &   90.4   &   2.1   &   1.0  &  0.3  \\  
RECX11 & 2.3~$\times$~10$^{3}$   &   6.2   &   4.6   &   90.9   &   2.9   &   1.7  &  0.6 \\  
RECX15 & 4.8~$\times$~10$^{3}$   &   6.5   &   2.8   &   96.7   &   0.3   &   0.3  &  0.1 \\  
RULUPI & 2.8~$\times$~10$^{4}$   &   7.2   &   17.6   &   79.8   &   2.0   &   0.7  &  0.3 \\  
SUAUR & 3.3~$\times$~10$^{4}$   &   7.3   &   8.4   &   88.3   &   2.1   &   1.2  &  0.4 \\  
TWHYA & 9.7~$\times$~10$^{3}$   &   6.8   &   16.7   &   71.9   &   6.5   &   4.9  &  0.9 \\  
UXTAU & 5.6~$\times$~10$^{3}$   &   6.5   &   4.8   &   92.2   &   1.7   &   1.3   &  0.3\\  
V4046SGR & 1.5~$\times$~10$^{4}$   &   7.0   &   5.3   &   91.8   &   1.2   &   1.8 &  0.9 
 \\  
\tableline
Average\tablenotemark{e}  &     &  7.0~$\pm$~0.5    &   8.4~$\pm$~5.2   &   88.1~$\pm$~7.3   &   2.1$~\pm$~1.6   &  1.4~$\pm$~1.2  &  0.5~$\pm$~0.4
\enddata
\tablenotetext{a}{Integrated 912~--~1650~\AA\ stellar+accretion FUV radiation field (not including molecular fluorescence lines, low-ionization atomic emission, or the ``1600~\AA\ Bump''), evaluated at 1 AU from the central pre-main sequence star.  } 
\tablenotetext{b}{Ratio of the integrated FUV radiation field at 1 AU to the average interstellar radiation field (1.6~$\times$~10$^{-3}$ erg cm$^{-2}$ s$^{-1}$; Habing 1968). } 
\tablenotetext{c}{Intrinsic Ly$\alpha$ emission, integrated over 1211~--~1221~\AA .} 
\tablenotetext{d}{Other stellar+accretion hot gas emission lines, \ion{C}{3} $\lambda$977~\AA\ + \ion{O}{6} $\lambda$$\lambda$1032,1038~\AA\ + \ion{N}{5} $\lambda$$\lambda$1239, 1243~\AA\ + \ion{He}{2} $\lambda$1640~\AA. } 
\tablenotetext{e}{Average quantities are calculated excluding DR Tau, whose Ly$\alpha$ profile reconstruction is compromised by multiple Ly$\alpha$ emission sources (\S A.1).} 
\end{deluxetable*}

\subsubsection{Ly$\alpha$ Line Emission}  The intrinsic Ly$\alpha$ flux (in units of erg cm$^{-2}$ s$^{-1}$), evaluated at 1 AU from the central star, is computed by integrating the reconstructed Ly$\alpha$ line profile from 1211~--~1221~\AA.  We note that we take the intrinsic reconstructed line profile as opposed to the line profile seen by the H$_{2}$ molecules at the disk surface after absorption by atomic hydrogen in the outflow (see Figure A.1).  A discussion of the Ly$\alpha$ profiles including the neutral outflow absorption component is presented by~\citet{schindhelm12b}, and the uncertainty on the intrinsic Ly$\alpha$ flux used here is $\sim$~10~--~30\%, depending on the detectability and S/N of the H$_{2}$ fluorescence lines used for the reconstruction~\citep{france12b}.  

The middle panels of Figure 5 show the correlation between Ly$\alpha$ and the FUV continuum ([$\rho$,$n$] = [0.71, 1.9~$\times$~10$^{-3}$]) and the \ion{C}{4} emission ([$\rho$,$n$] = [0.77, 4.8~$\times$~10$^{-4}$]).  The Ly$\alpha$ flux is observed to correlate with both quantities, but with less significance than the correlations between \ion{C}{4} and \ion{N}{5}, and \ion{C}{4} and the FUV continuum.  This is possibly confirmed by another representation on the bottom panels of Figure 5, where the ratio of Ly$\alpha$/FUV continuum is plotted against the published mass accretion rates ($left$, [$\rho$,$n$] = [$-$0.31, 0.24]) and the \ion{C}{4} fluxes ($right$, [$\rho$,$n$] = [$-$0.50, 0.047]).  The plots are qualitatively similar; Figure 5 ({\it bottom left}) does not show a significant anti-correlation between Ly$\alpha$/FUV continuum and the mass-accretion rate, Ly$\alpha$/FUV continuum and \ion{C}{4} show a weak anti-correlation suggesting that the FUV continuum is more tightly correlated with the accretion than the Ly$\alpha$ emission.  

\begin{figure}
\figurenum{7}
\begin{center}
\epsfig{figure=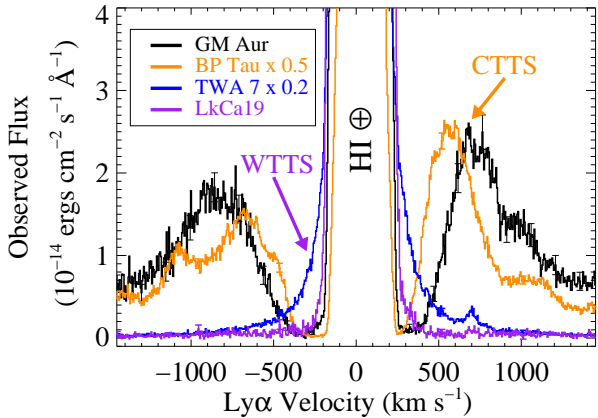,width=3.5in,angle=00}
\vspace{+0.0in}
\caption{ A comparison between observed CTTS and WTTS Ly$\alpha$ profiles for a few example objects.  The CTTS Ly$\alpha$ profiles (GM Aur and BP Tau) uniformly display higher fluxes and line widths ($\Delta$$v$~$>$~$\pm$~500 km s$^{-1}$) than their WTTS counterparts (TWA 7 and LkCa19), suggesting that accretion processes contribute (and probably dominate) to the total Ly$\alpha$ luminosity.  
The narrow central emission is from Earth's geocorona and the broad depression seen at the core of the CTTS line profiles is due to resonant scattering in the interstellar medium along the line-of-sight.
\label{cosovly}}
\end{center}
\end{figure}
The production of Ly$\alpha$ is still likely dominated by accretion processes.  Figure 7 shows the observed Ly$\alpha$ emission lines from two representative CTTSs from this work (GM Aur and BP Tau) and two WTTSs (TWA 7 and LkCa19).  In all cases, the CTTS Ly$\alpha$ lines are much stronger and broader than WTTS Ly$\alpha$ profiles for a similar age and stellar mass.  All CTTSs display broad Ly$\alpha$ wings to $v_{Ly\alpha}$~$>$~500 km s$^{-1}$, and many show Ly$\alpha$ emission wings in excess of $\pm$~1000 km s$^{-1}$.  
There may be several physical reasons that the correlation between Ly$\alpha$ and \ion{C}{4} is weaker than between the FUV continuum and \ion{C}{4}, including uncertainties in the Ly$\alpha$ reconstruction process, particularly when there are multiple sources of Ly$\alpha$ photons in the systems (stellar atmosphere, outflows and/or disk surface) and resonant scattering of Ly$\alpha$ photons in the gas-rich CTTS circumstellar environment.  

 \subsection{Absolute Fluxes at Planet-forming Radii and Relative Contributions} 

In order to understand the UV-driven photochemistry of the protoplanetary environment, one needs constraints on both the absolute flux and the shape of the illuminating radiation field.  The absolute flux of the radiation environment determines the photoexcitation and photodissociation rates for disk molecules and the SED determines the primary scattering and absorption mechanisms for diffusing FUV photons through the disk~\citep{bergin03,bethell11,fogel11}.  In Table 2, we give the total, reddening-corrected flux from our targets (evaluated at 1 AU to enable inter-comparison) integrated over the 912~--~1650~\AA\ bandpass.  In the second column, this quantity is given in units of the average interstellar radiation field from~\citet{habing68}.  The average incident flux from the accreting central star is~$\sim$~10$^{7}$ times the contribution from scattered OB starlight at 1 AU from the central star, although the shapes of the CTTS radiation field and the interstellar field are very different.  

By spectrally resolving the various components of the FUV field and including the reconstructed Ly$\alpha$ emission, we can provide an inventory of the FUV luminosity sources in $\sim$~1~--~10 Myr CTTSs.  Table 2 shows the relative contribution of each component to the total FUV irradiance from each target.  As argued by previous authors~\citep{bergin03,herczeg04,schindhelm12b}, Ly$\alpha$ dominates the FUV radiation output from these sources, with an average fractional luminosity of 88.1~$\pm$~7.3~\%\footnote{The only object for which Ly$\alpha$ may not be the largest contributor to the FUV radiation field is DR Tau. However, the Ly$\alpha$ reconstruction for this star is rather uncertain due to the presence of multiple sources of Ly$\alpha$ emission in the system, most likely a second, narrow component associated with an outflow. }. 
The FUV continuum is the second largest energy source, with 8.4~$\pm$~5.2~\%.  The contribution from the \ion{C}{4} $\lambda$$\lambda$1548,1550~\AA\ doublet is 2.1~$\pm$~1.6~\%, and the remaining hot gas emission lines make up $<$~5\% in all cases, although the exact contribution is rather uncertain because we assumed the TW Hya \ion{C}{3} and \ion{O}{6} fluxes, scaled by the relative \ion{C}{4} fluxes, for the remaining targets.  Lower ionization species likely contribute $<$ 1\% to the total field strength (see A.2).  

The primary transitions leading to dissociation of H$_{2}$ and CO, through the H$_{2}$ $B$~--~$X$ and $C$~--~$X$ band systems (radiatively dissociative) and the CO $E$~--~$X$ band system (predissociative), are located at $\lambda$~$\leq$~1110~\AA.  The $\lambda$~$\leq$~1110~\AA\ radiation field also plays an important role in the production of photo-electrons that heat the gas in the disk surface layers~\citep{pedersen11}.  The 912~--~1110~\AA\ bandpass contains between 0.1~\% and 1.8~\% of the total FUV flux, with an average (excluding DR Tau) of 0.5~$\pm$~0.4~\%.   On average, the 912~--~1110~\AA\ CTTS radiation field has a flux of $\sim$~150 erg cm$^{-2}$ s$^{-1}$ at 1 AU from the central star.  


\subsection{Comparison with Low-resolution Spectra}

In this subsection, we quantify the impact of low-resolution data on the determination of the total FUV flux from CTTSs.  Previous large FUV spectroscopic samples of CTTSs were acquired at low spectral resolution ($R$~$\sim$~80~--~1000; Johns-Krull et al. 2000; Valenti et al. 2000; Ingleby et al. 2009, 2011; Yang et al. 2012).  Most of these observations did not have access to Ly$\alpha$ directly due to instrumental bandpass limitations and/or contamination by geocoronal emission.  Furthermore, low-resolution prevents the necessary measurements of individual H$_{2}$ or CO fluorescent emission lines necessary for reconstructing the Ly$\alpha$ emission profile.  Consequently, most quoted measurements of the FUV radiation field strength are integrated over $\approx$ 1230~--~1700~\AA~\citep{ingleby11,yang12}, $excluding$ the dominant FUV flux source, \ion{H}{1} Ly$\alpha$.  Low resolution data are also insufficient for separating the accretion-powered FUV continuum from the wealth of narrow molecular emission lines and the molecular continuum (the ``1600~\AA\ Bump''). In order to quantify the differences between high- and low-resolution CTTS FUV radiation fields, data for 10 objects with previous low-resolution FUV observations from $HST$ were retrieved from the~\citet{yang12} atlas of low-resolution T Tauri star spectra hosted at MAST.  

The average flux ratio between the observed low-resolution 1235~--~1700~\AA\ flux from the~\citet{yang12} atlas ($F(1235-1700)_{LR}$) and the 1235~--~1700~\AA\ fluxes from this work ($F(1235-1700)_{fit}$), $F(1235-1700)_{LR}$/$F(1235-1700)_{fit}$, is 1.41~$\pm$~0.52 (Table 3).  The average being greater than unity most likely reflects the inclusion of Ly$\alpha$-pumped H$_{2}$ and CO fluorescence in low-resolution data and the scatter is most likely the time-variability that is characteristic of these sources (recall these data were acquired over a multi-year baseline; and see e.g., Giovannelli et al. 1995; G{\'o}mez de Castro \& Fern{\'a}ndez 1996).  The two sources which display the largest FUV variability are DM Tau and SU Aur, showing a factor of $\sim$~3 variability between the two epochs.  If we screen sources farther than 3--$\sigma$ from the mean of the distribution,  the average drops to $F(1235-1700)_{LR}$/$F(1235-1700)_{fit}$ = 1.06~$\pm$~0.34.

\begin{deluxetable}{lccc}
\tabletypesize{\normalsize}
\tablecaption{Comparison to Low-Resolution CTTS Spectra \label{lya_lines}}
\tablewidth{0pt}
\tablehead{ 
\colhead{Target} & \colhead{$HST$ Mode\tablenotemark{a}} &   
\colhead{ $\frac{ F(1235-1700)_{LR} }{ F(1235-1700)_{fit} }$ \tablenotemark{b} }   & 
\colhead{ $\frac{ F(1235-1700)_{LR} }{ F(1150-1700)_{fit} }$ \tablenotemark{c} } }
\startdata
AATAU & ACS   &   1.18   &   0.07   \\  
BPTAU & STIS   &   0.70   &   0.21     \\  
DETAU & ACS   &   1.68   &   0.27   \\  
DMTAU & STIS   &   2.73   &   0.31   \\  
DRTAU & ACS   &   0.77   &   0.55    \\  
GMAUR & ACS   &   0.99   &   0.14     \\  
GMAUR & STIS   &   0.96  &   0.14     \\  
HNTAU & ACS   &   1.04   &   0.17    \\  
LKCA15 & STIS   &   1.43   &   0.19     \\  
SUAUR & STIS  &   3.24   &   0.60   \\  
V4046SGR & ACS   &   0.76   &   0.05   \\
\tableline
Average   &     &  1.41~$\pm$~0.52    &   0.25~$\pm$~0.17   
\enddata
\tablenotetext{a}{$HST$ Low-resolution mode used for comparison. Low-resolution data taken from~\citet{yang12}.   ACS refers to ACS/SBC PR130L mode, STIS refers to STIS G140L. } 
\tablenotetext{b}{Ratio of the observed low-resolution spectrum from 1235~--~1700~\AA\ to the extracted, intrinsic CTTS spectrum from 1235~--~1700~\AA. } 
\tablenotetext{c}{Ratio of the observed low-resolution spectrum from 1235~--~1700~\AA\ to the extracted, intrinsic CTTS spectrum from 1150~--~1700~\AA, including the reconstructed Ly$\alpha$ emission line.  
}
\end{deluxetable}

Comparing the ratio of the observed low-resolution 1235~--~1700~\AA\ flux from the~\citet{yang12} atlas and our 1150~--~1700~\AA\ fluxes that {\it include the reconstructed Ly$\alpha$ profiles} ($F(1150-1700)_{fit}$), we find  $F(1235-1700)_{LR}$/$F(1150-1700)_{fit}$~=~0.25~$\pm$~0.17.  Applying the 3--$\sigma$ screen, $F(1235-1700)_{LR}$/$F(1150-1700)_{fit}$~=~0.17~$\pm$~0.08.  The general result is that even with the additional contribution from H$_{2}$ emission, and the uncertainty associated with the time-variability of the sources, low-resolution FUV spectra that do not include a reconstructed Ly$\alpha$ emission line underestimate the total FUV flux by a factor of $\approx$~6.  While a factor of $\approx$~6 is the average flux underestimate, Table 3 shows that in the extreme cases (e.g., AA Tau and V4046 Sgr), the low-resolution 1235~--~1700~\AA\ data underestimate the true intrinsic stellar+accretion FUV radiation field strength by factors of ~$\geq$~15.  

Our conclusion from this comparison with low-resolution data in the literature is that ACS/SBC PR130L and STIS G140L-based flux estimates underestimate the true FUV radiation field strength from CTTSs by approximately an order-of-magnitude.  Our findings quantitatively support the assertions of previous authors~\citep{bergin03,herczeg04,schindhelm12b} that Ly$\alpha$ must be included to make an accurate of estimate of $both$ the strength and the shape of the FUV radiation field in protoplanetary environments around low-mass stars.  It should also be emphasized that the largest uncertainty on the inferred local FUV radiation field strength is likely the uncertainty on the line-of-sight reddening, both interstellar and circumstellar.  Typical dispersions on the derived values of $A_{V}$ in the literature are 0.3~--~1.0 magnitudes, the relative contributions of interstellar and circumstellar grains are unclear~(see e.g., McJunkin et al. 2014 and references therein), and the scattering and extinction properties of grains in regions like Taurus may be quite different than those found in the diffuse interstellar medium~\citep{calvet04}.\nocite{mcjunkin14}  With empirically-derived UV radiation fields now available, the uncertainty on the extinction is likely the limiting factor on our knowledge of the absolute flux of the energetic radiation environments around young stars.

\section{Summary}

We have presented new analyses of the 1150~--~1700~\AA\ spectra of 16 Classical T Tauri Stars observed with the {\it Hubble Space Telescope}.   We have used the combination of the high spectral resolution and high S/N of these data to create the first high-resolution intrinsic stellar+accretion FUV (912~--~1700~\AA) radiation fields from these sources.   The radiation fields include a detailed extraction of and fit to the true FUV continuum, the reconstructed Ly$\alpha$ emission line profile, and empirically determined profiles of hot gas emission lines (e.g., \ion{N}{5}, \ion{C}{4}, \ion{He}{2},  etc.) from which molecular emission from the circumstellar disk has been removed.  These radiation fields are publically available in a machine-readable format at {\tt http://cos.colorado.edu/$\sim$kevinf/ctts\_fuvfield.html} for use in disk chemistry or disk evolution models.  

Using \ion{C}{4} line fluxes as a proxy for magnetospheric accretion, we demonstrate that the Ly$\alpha$ emission lines and the FUV continua are likely generated, at least in part, by accretion.  We find a spectral break between the NUV and FUV continua suggesting that they arise in the different regions near the stellar surface.   The broad line profiles of Ly$\alpha$ ($\Delta$$v$~$>$~$\pm$~500 km s$^{-1}$) in the CTTSs, relative to non-accreting WTTS profiles, also suggest an accretion origin for this emission.  We find that the flux of the typical Classical T Tauri Star FUV radiation field at 1 AU from the central star is $\sim$~10$^{7}$ times the average interstellar radiation field.   The Ly$\alpha$ emission line contributes an average of 88\% of the total FUV flux, with the FUV continuum accounting for an average of 8\%.  FUV radiation fields that do not include the contribution from Ly$\alpha$ underestimate the total FUV flux by factors of 2~--~15.  This can have an impact on the evolution of protoplanetary disks and the disk chemistry of species with excitation and dissociation transitions coincident with the Ly$\alpha$ emission line.   

\acknowledgments
  This work received support from NASA grant NNX08AC146 to the University of Colorado at Boulder ($HST$ programs 11533 and 12036) and made use of data from $HST$ GO programs 8041 and 11616.  KF acknowledges support from a Nancy Grace Roman Fellowship, and thanks Joanna Brown, Ewine van Dishoeck, Simon Bruderer, and Greg Herczeg for enjoyable discussion relating to FUV radiation fields and disk photochemistry.  We thank the anonymous referee for several constructive suggestions for improving the description of the FUV continuum.  
KF also appreciates the hospitality of White Sands Missile Range, where a portion of this work was carried out.  

\appendix

\section{Assembly of High-Resolution Ultraviolet Radiation Fields} 

In this Appendix, we describe the details of the Ly$\alpha$ reconstruction (A.1), the removal of H$_{2}$ emission from hot gas lines (A.2), and the measurement of the FUV continuum (A.3) for the 16 targets described herein.  The reddening corrected, individual spectral components are displayed in the accompanying figures (Figures A.1~--~A.6), scaled to the flux (in units of erg cm$^{-2}$ s$^{-1}$ \AA$^{-1}$) at 1 AU in order to present a comparison that is (mostly) independent of distance and stellar radius.  The total, high-resolution FUV radiation fields from these 16 stars, and the individual component spectra are available in machine-readable format at the following website: {\tt http://cos.colorado.edu/$\sim$kevinf/ctts\_fuvfield.html}.

\begin{figure}
\figurenum{A.1}
\begin{center}
\epsfig{figure=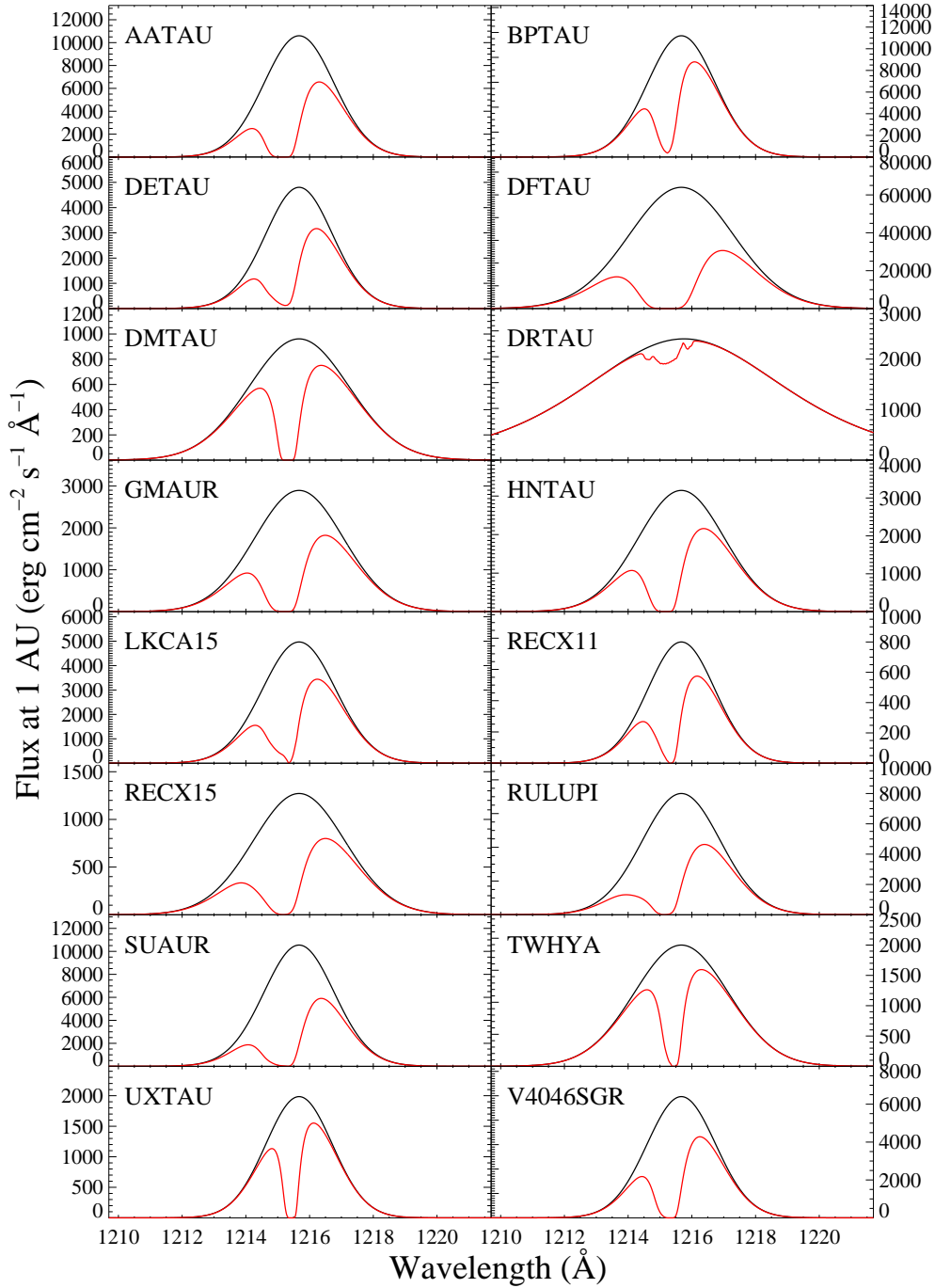,width=4.5in,angle=00}
\vspace{+0.45in}
\caption{Reconstructed Ly$\alpha$ emission line profiles.  The black curves are the average of intrinsic Gaussian emission line profile distribution that is consistent with the fluorescent H$_{2}$ emission observed in these systems~\citep{schindhelm12b}.    The red curves are the average Ly$\alpha$ profiles as observed at the disk surface, the self-reversal of the line profile is caused by a blueshifted ($\sim$~$-$100 km s$^{-1}$) atomic wind that is observed for all stars.  The reconstructions of DF Tau and DR Tau are new for this work, although a single emission component does not provide a good fit to the Ly$\alpha$ profile of DR Tau (see \S A.1).  
\label{cosovly}}
\end{center}
\end{figure}

\subsection{Ly$\alpha$ Reconstruction}

The \ion{H}{1} Ly$\alpha$ emission line dominates the energy output of CTTSs in the FUV bandpass, contributing $\sim$~70~--~90\% to the total 912~--~1700~\AA\ flux~\citep{herczeg04,schindhelm12b}.  This line, produced in the protostellar atmosphere, accretion shocks, and extended outflows~\citep{walter03} can be hundreds of km s$^{-1}$ wide~\citep{schindhelm12b}, with luminosities ranging from roughly 0.25~--~40\%~$L_{\odot}$.  Due
to resonant scattering of neutral hydrogen in the interstellar and circumstellar media, the intrinsic Ly$\alpha$ radiation field cannot be directly measured for any star other than the Sun.   In order to produce an accurate estimate of the local Ly$\alpha$ flux incident on
the disk surface, the emission lines must be reconstructed using additional observational constraints.   \citet{wood02} and \citet{wood04} pioneered a technique of Ly$\alpha$ emission profile reconstruction using the fluorescent H$_{2}$ lines as a tracer of the Ly$\alpha$ radiation field incident on  the disk surface.  This method, adapted to CTTS disks by~\citet{herczeg04}, uses the measured H$_{2}$ emission line strengths for a given [$v^{'}$,$J^{'}$]~$\rightarrow$~[$v^{''}$,$J^{''}$] progression to determine the total absorbed Ly$\alpha$ flux at a given pumping wavelength.  The grid of Ly$\alpha$ fluxes is then fit with an intrinsic line shape and intervening absorber to infer the local Ly$\alpha$ profile and reproduce the observed molecular fluorescence spectra.  We have recently extended this technique to model the Ly$\alpha$ radiation fields and FUV CO emission from 7 stars~\citep{schindhelm12a}.

With the advent of larger medium-resolution CTTS samples observed with the $HST$-COS, the fluorescent H$_{2}$-based Ly$\alpha$ reconstruction can be applied to larger target samples.  In the present work, we adopt the reconstructed Ly$\alpha$ radiation fields from~\citet{schindhelm12b} as the input to our intrinsic CTTS FUV spectra.  
They use the fluxes from 12 fluorescent H$_{2}$ progressions~\citep{france12b} to infer the intrinsic Ly$\alpha$ radiation fields, assuming a single component Gaussian Ly$\alpha$ emission line, a blue-shifted \ion{H}{1} outflow component, and uniform covering fraction for the absorbing H$_{2}$.  
In addition to the 14 targets that were presented in the Schindhelm et al. sample, we have added new Ly$\alpha$ line reconstructions of DF Tau and DR Tau.    DR Tau could not easily be fit by our single-component model, with enhanced fluorescent emission pumped near the Ly$\alpha$ line core.  We interpret this as a second source of Ly$\alpha$ photons, most likely produced in an outflow, as has been observed in T Tau~\citep{walter03,saucedo03}.   The Ly$\alpha$ spectra are presented in Figure A.1.  The intrinsic profiles (normalized to the flux at 1 AU) are shown in black and the outflow-absorbed profiles as observed at the disk surface are shown in red.

\begin{figure*}
\figurenum{A.2}
\begin{center}
\epsfig{figure=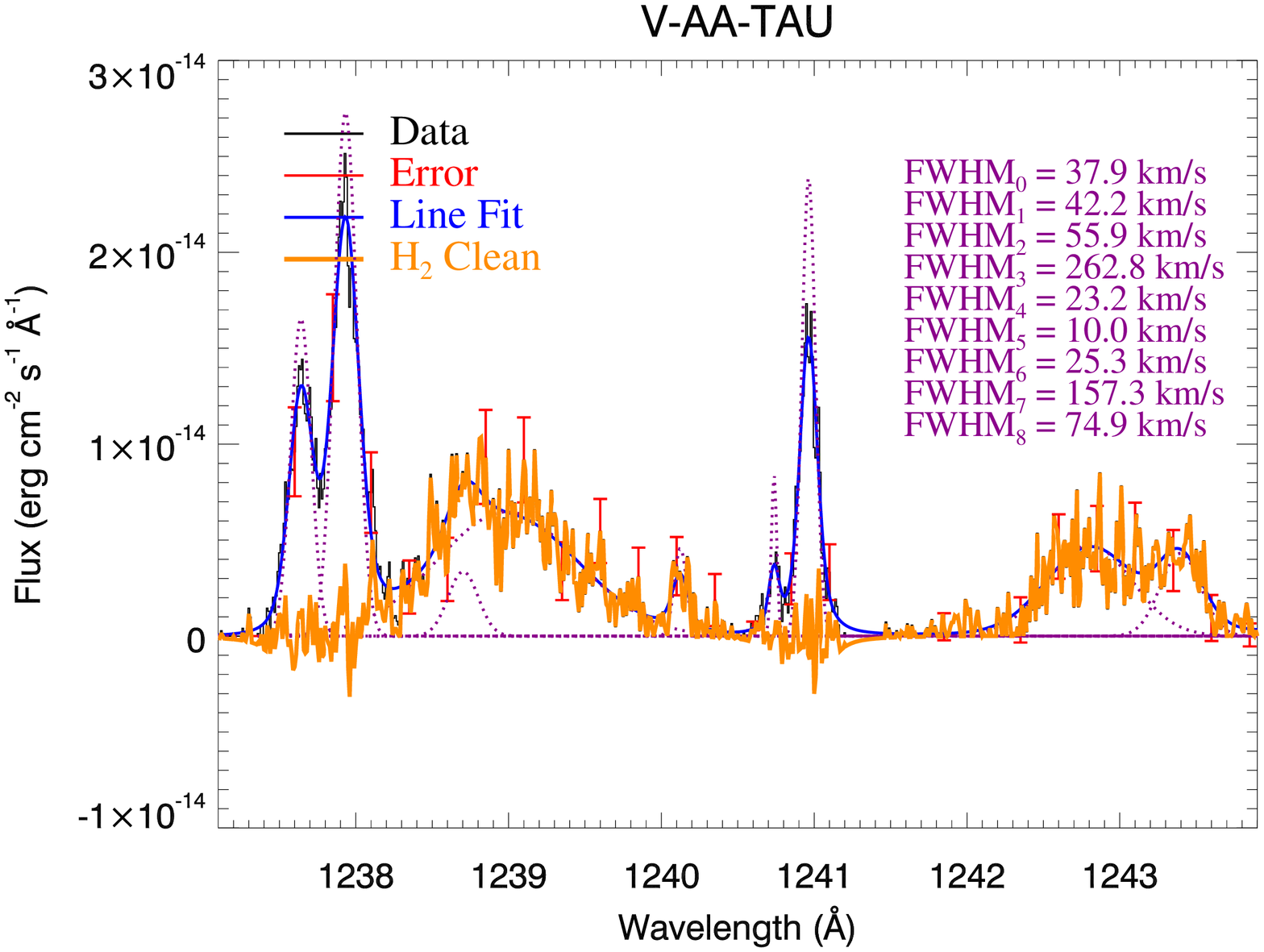,width=2.5in,angle=90}\hspace{-0.34in}
\epsfig{figure=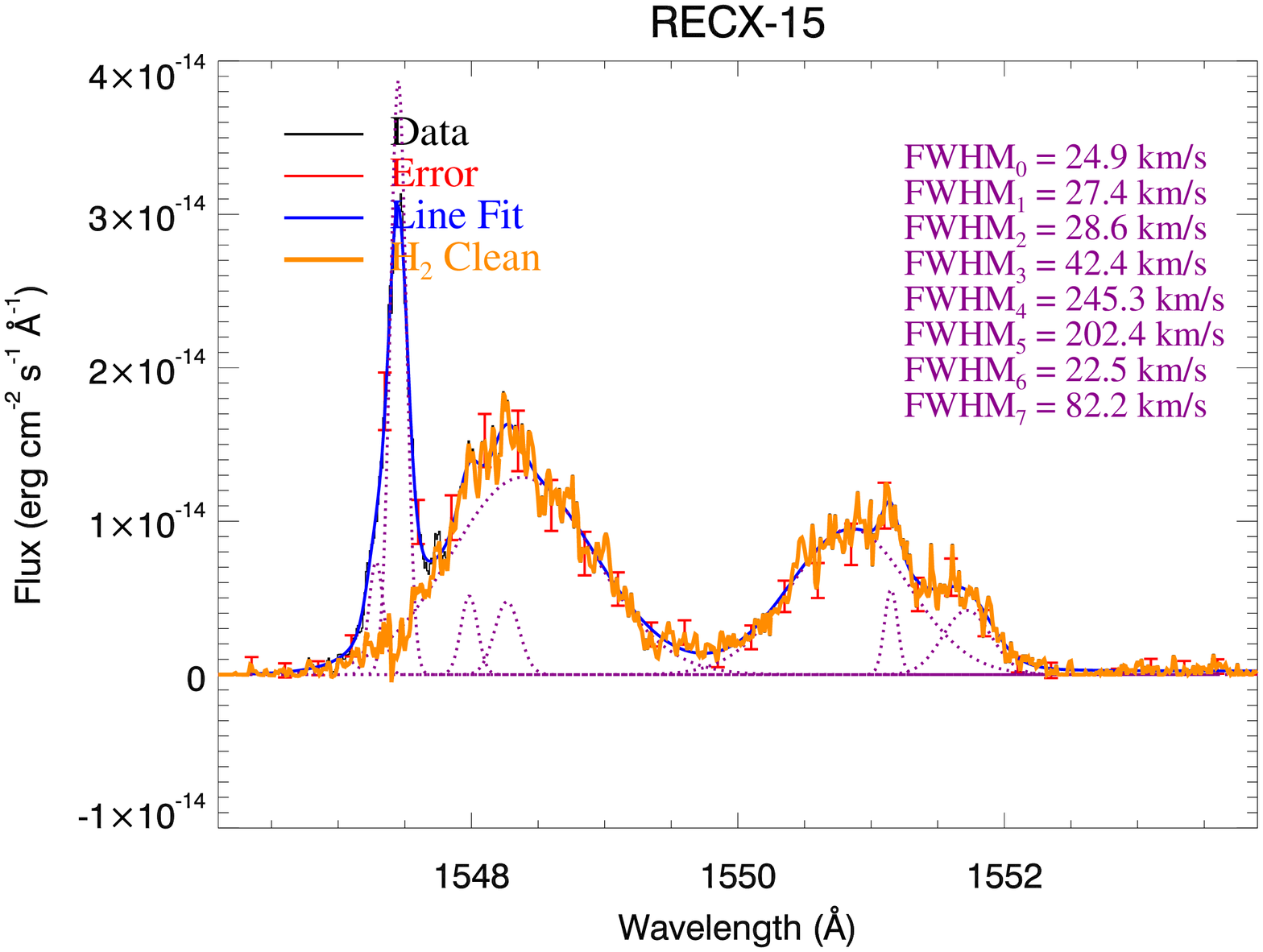,width=2.5in,angle=90}
\epsfig{figure=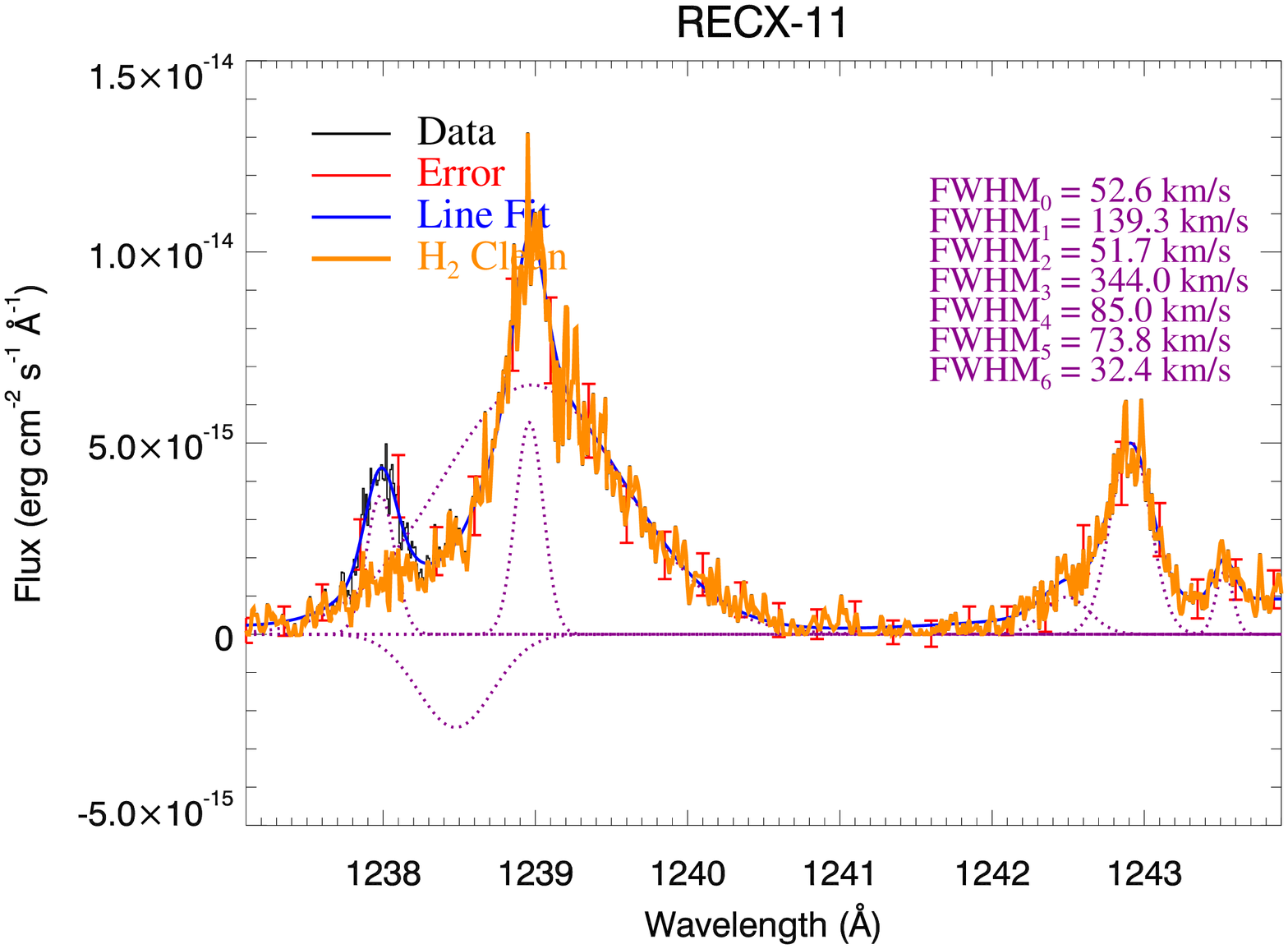,width=2.5in,angle=90}\hspace{-0.34in}
\epsfig{figure=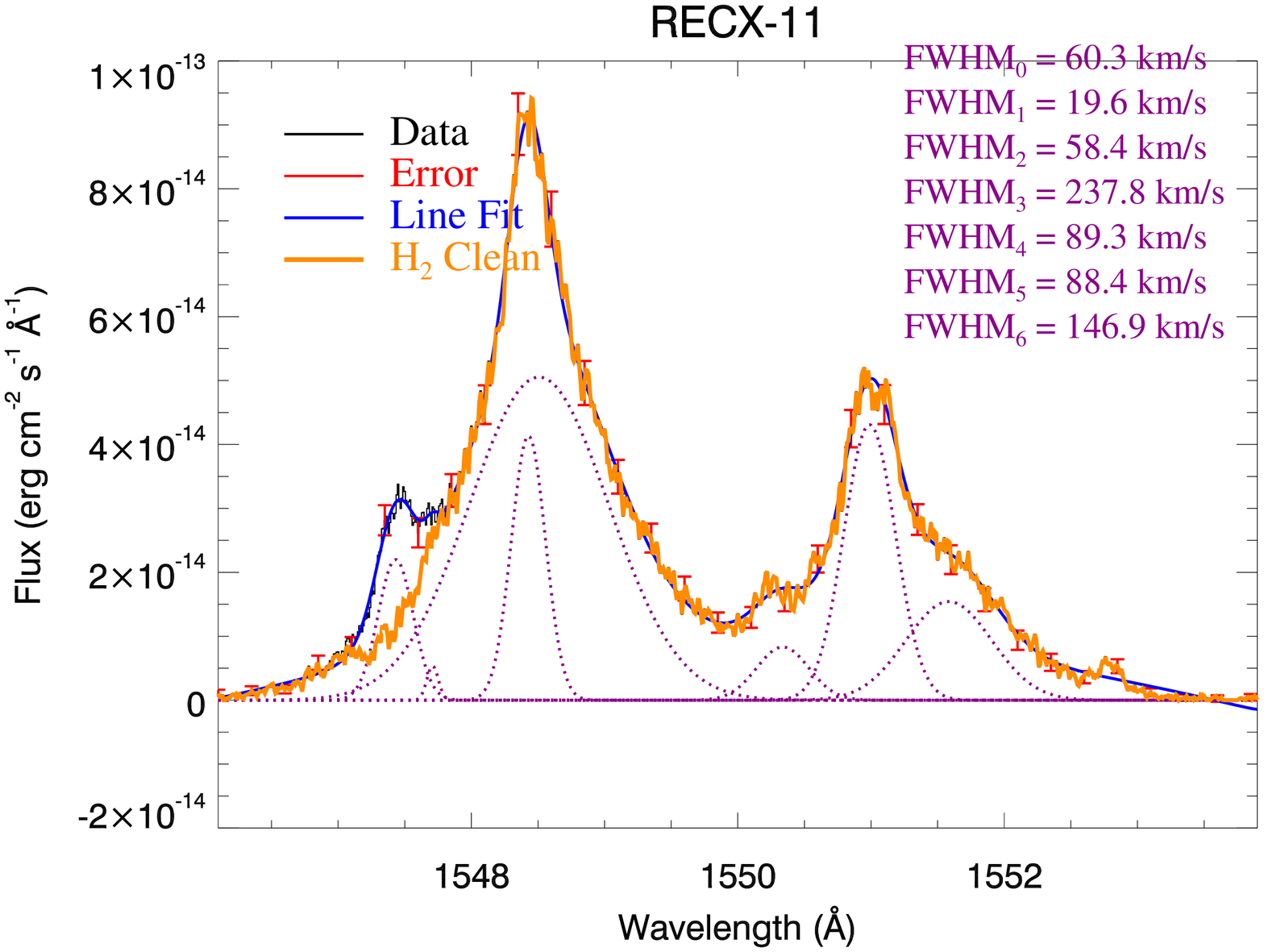,width=2.5in,angle=90}
\vspace{+0.45in}
\caption{Examples of the H$_{2}$ removal process for \ion{N}{5} ($left$) and \ion{C}{4} ($right$).  The examples shown here are representative of the molecular ``cleaning'' performed for all 16 targets; cases with large H$_{2}$ contamination (AA Tau and RECX-15; $top$) and cases with minimal H$_{2}$ contamination (RECX-11; $bottom$).   
The continuum-subtracted line profiles are shown as the thin black line (representative error bars displayed in red), the multi-Gaussian parameterization is shown as the solid blue line (individual components are dotted magenta lines; FWHMs of each component listed in legend), and the post-cleaning hot gas line profiles used in the radiation field creation are shown as the thick orange curves.  
\label{cosovly}}
\end{center}
\end{figure*}

\subsection{Hot Gas Emission Lines and Subtraction of H$_{2}$ Fluorescent Emission Lines}

The accretion shocks (including pre- and post-shock regions and the heated stellar photosphere near the base of the accretion columns) and stellar chromosphere/transition regions will produce other strong emission lines formed
at temperatures from $T_{form}$~$\sim$~10$^{4}$~--~3~$\times$~10$^{5}$ K.  The higher ionization lines are above the typical temperatures of the intervening interstellar material, and therefore are not subject to strong attenuation before reaching Earth.   For hot gas lines formed near the stellar surface, we use the observed emission spectra, subtracted for a local continuum level, and then combined with the FUV continuum fit described below.  The emission lines included in our FUV radiation field are \ion{C}{3} $\lambda$977, \ion{O}{6} $\lambda$$\lambda$1032,1038, \ion{H}{1} Ly$\alpha$ $\lambda$1216 (described above), \ion{N}{5} $\lambda$$\lambda$1239,1243, \ion{C}{4} $\lambda$$\lambda$1548,1550, and \ion{He}{2} $\lambda$1640~\AA.  We do not include emission lines of silicon because these lines are weak and occasionally depleted in CTTS accretion spectra~\citep{herczeg02,ardila13}.  Inclusion of the \ion{C}{3} and \ion{O}{6} lines is important because hot gas lines will contribute a significant fraction of the stellar+shock flux at $\lambda$~$<$~1100~\AA\ where most of the H$_{2}$ and CO absorption bands reside.  However, only TW Hya, DF Tau, RU Lupi, and V4046 Sgr have moderate S/N spectra in the {\it Far-Ultraviolet Spectroscopic Explorer (FUSE)} archive, owing to the faintness of these sources at the shortest UV wavelengths.  In order to present a uniform analysis, we take the emission profiles of the highest quality sub-1100~\AA\ data (TW Hya) as representative of all of the CTTS spectra.  The TW Hya \ion{C}{3} and \ion{O}{6} emission line profiles are then scaled to the \ion{C}{4} flux level for each source.  This may introduce substantial errors to the sub-1100 \AA\ data, but in the absence of a self-consistent model for the entire FUV bandpass, this simple prescription was adopted.

\begin{figure*}
\figurenum{A.3a}
\begin{center}
\epsfig{figure=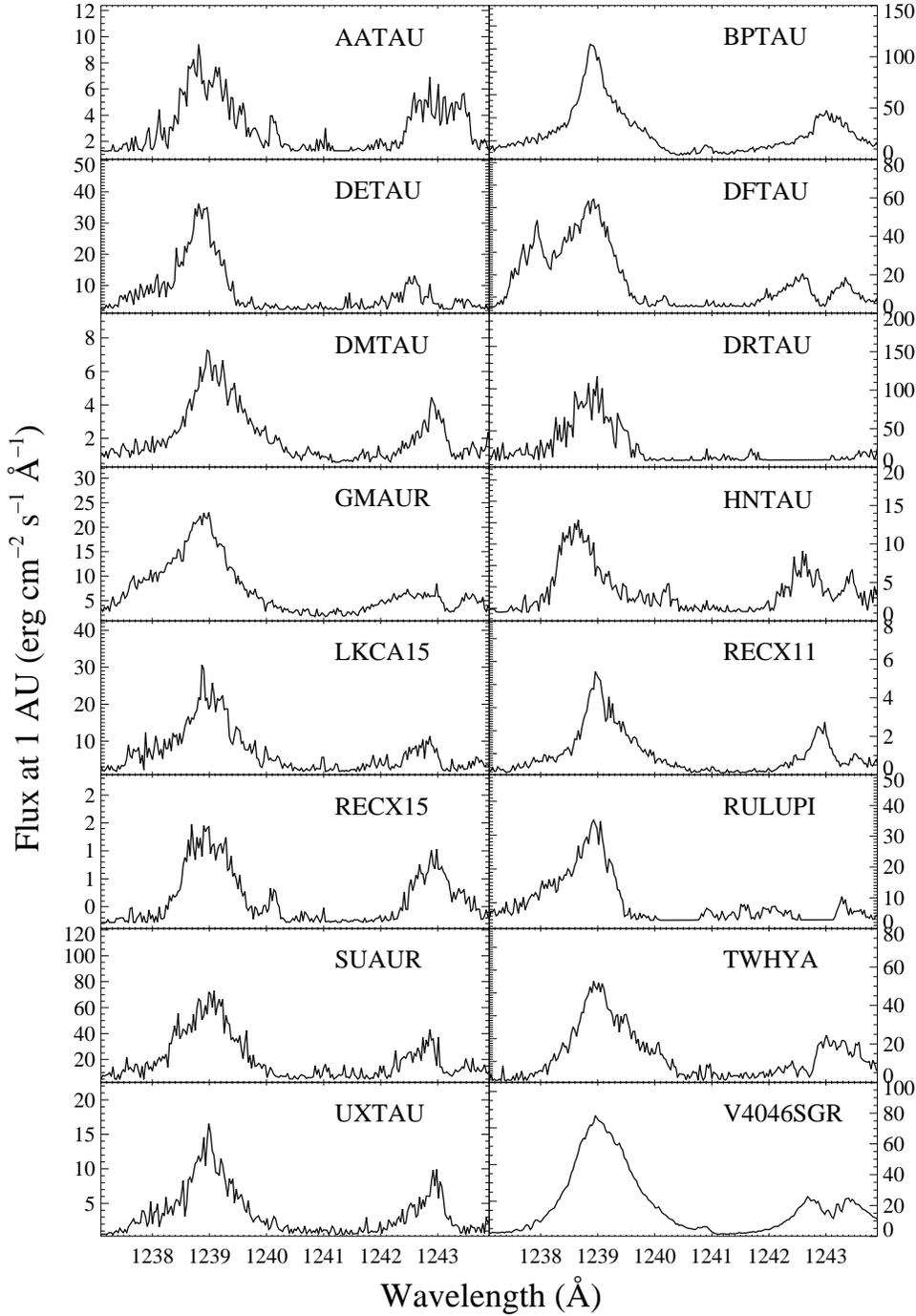,width=4.5in,angle=00}
\vspace{+0.45in}
\caption{Observed \ion{N}{5} profiles after subtracting the fluorescent H$_{2}$ emission lines.  
\label{cosovly}}
\end{center}
\end{figure*}

\begin{figure}
\figurenum{A.3b}
\begin{center}
\epsfig{figure=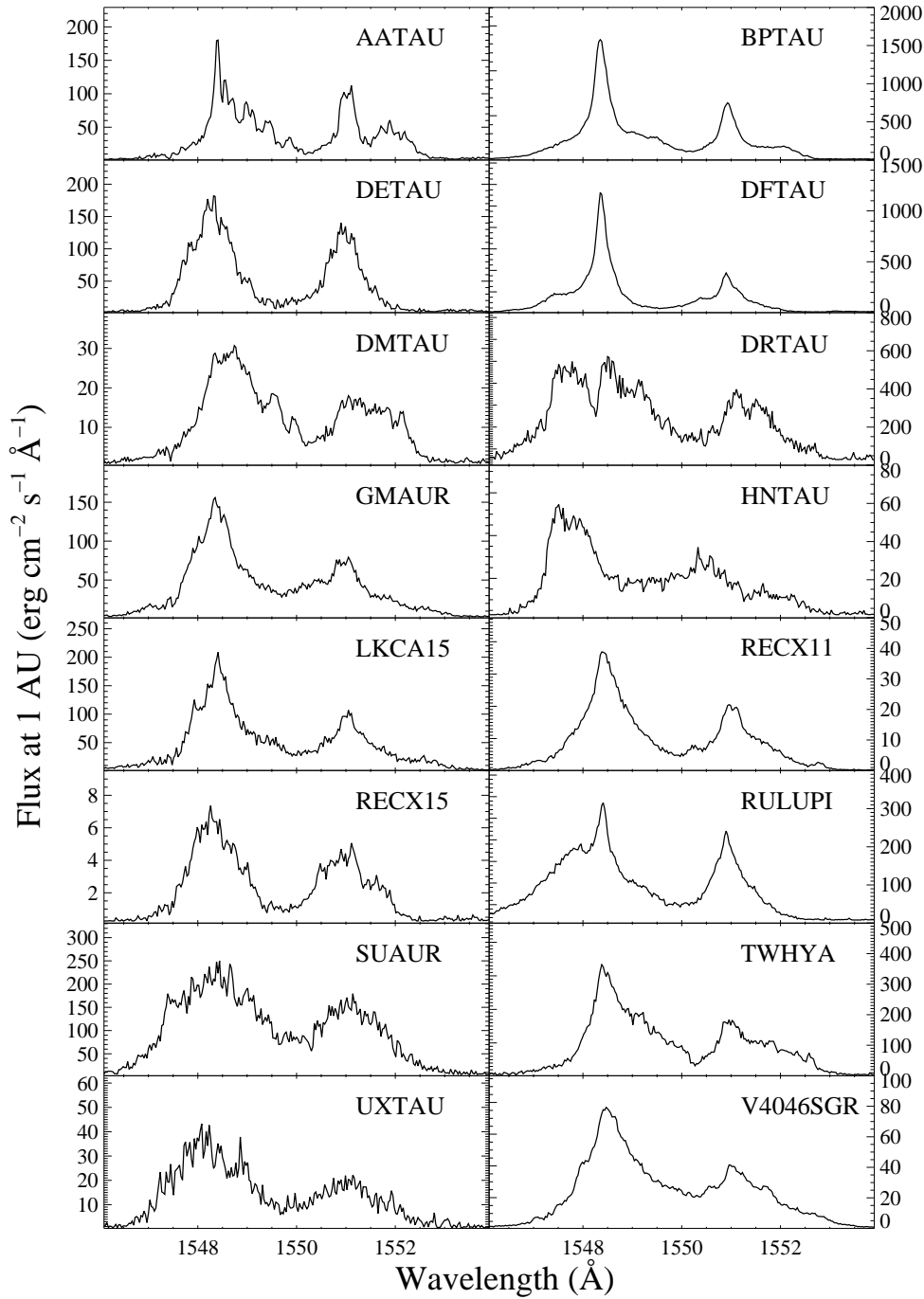,width=4.5in,angle=00}
\vspace{+0.45in}
\caption{Observed \ion{C}{4} profiles after subtracting the fluorescent H$_{2}$ emission lines.  A detailed discussion of the origin of these lines is presented by~\citet{ardila13}.  
\label{cosovly}}
\end{center}
\end{figure}

\begin{figure}
\figurenum{A.3c}
\begin{center}
\epsfig{figure=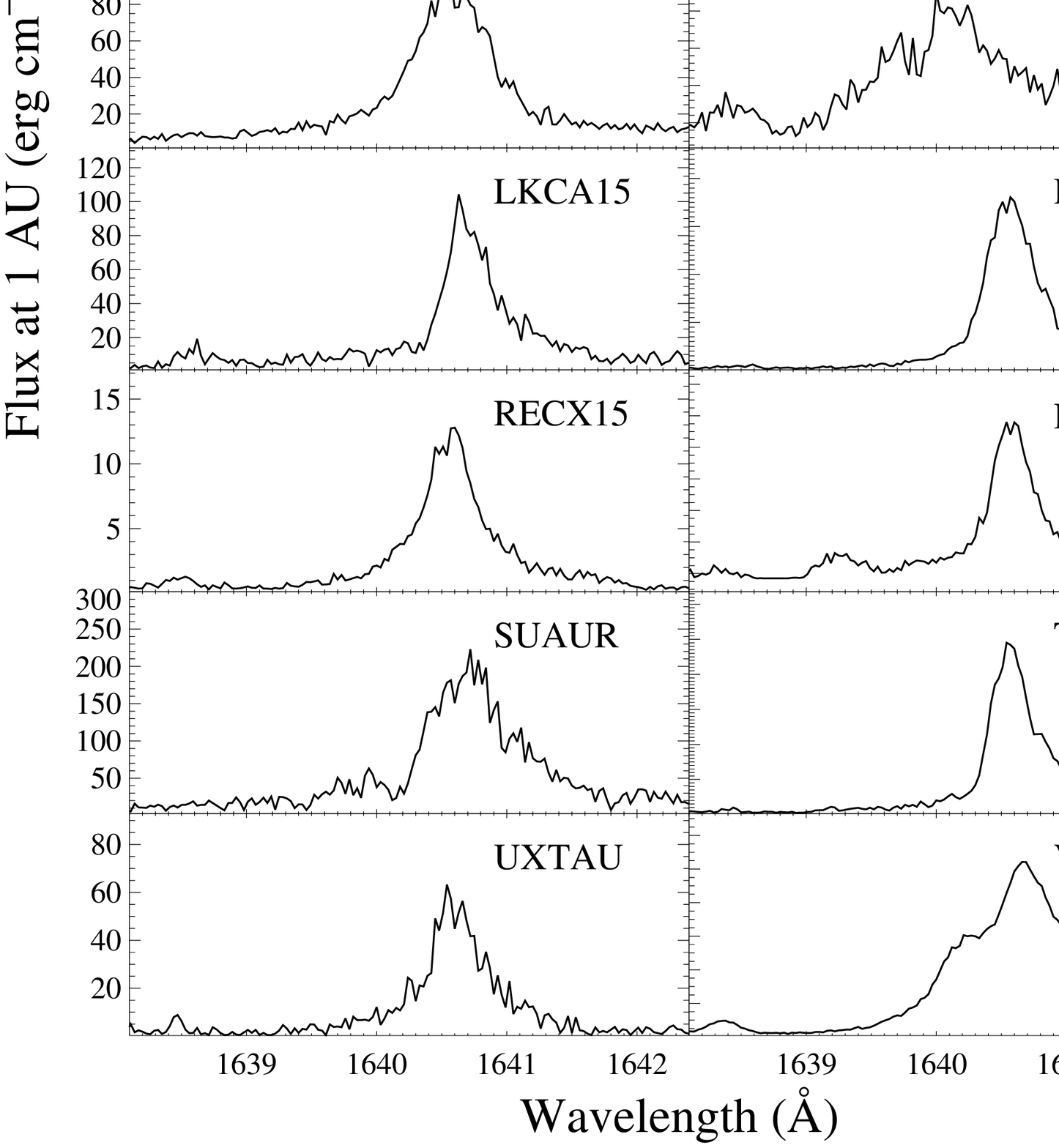,width=4.5in,angle=00}
\vspace{+0.45in}
\caption{Observed \ion{He}{2} profiles. 
\label{cosovly}}
\end{center}
\end{figure}

Emission from \ion{N}{5} and \ion{C}{4} are blended with several fluorescent H$_{2}$ emission lines; in order to present the intrinsic hot gas lines without contamination from the surrounding circumstellar material, we have developed a technique to fit and remove these lines.  For \ion{N}{5}, we load the 1237.1~--~1243.9~\AA\ spectra into the multi-Gaussian fitting routine described in~\citet{france12b} and fit all of the observed features with Gaussian emission lines convolved with the appropriate wavelength dependent $HST$+COS line-spread function~\citep{kriss11}\footnote{The COS LSF experiences a wavelength dependent non-Gaussianity due to the introduction of mid-frequency wave-front errors produced by the polishing errors on the $HST$ primary and secondary mirrors; {\tt http://www.stsci.edu/hst/cos/documents/isrs/}.}.
Emission lines with FWHM $<$~100 km s$^{-1}$ and within $\pm$~40 km s$^{-1}$ of the rest wavelength of a known H$_{2}$ emission line pumped by Ly$\alpha$ [$B$~--~$X$ (2~--~2)$R$(11) $\lambda$1237.54~\AA, $B$~--~$X$ (1~--~2)$P$(8) $\lambda$1237.87~\AA, and $C$~--~$X$ (1~--~5)$R$(9) $\lambda$1240.87~\AA] were then subtracted from the data.  For \ion{C}{4}, the 1546.1~--~1553.9~\AA\ spectra were fitted and H$_{2}$ lines [$B$~--~$X$ (1~--~8)$R$(3) $\lambda$1547.34~\AA\ and $B$~--~$X$ (3~--~7)$P$(17) $\lambda$1551.76~\AA] subtracted.  
The ``H$_{2}$ cleaning'' process is displayed graphically in Figure A.2.  The final \ion{N}{5}, \ion{C}{4}, and \ion{He}{2} profiles used in the FUV radiation fields, scaled to the flux at 1 AU, are displayed in Figures A.3($a$~--~$c$), respectively.

We compared the ``H$_{2}$ cleaned'' \ion{N}{5} $\lambda$1238 and \ion{C}{4} $\lambda$1548 + $\lambda$1550 raw line fluxes (without correction for reddening or scaling to 1 AU) to the fluxes for these lines presented by~\citet{ardila13}.  We find that our \ion{C}{4} fluxes agree to better than 10\% and the \ion{N}{5} fluxes derived in this work agree to within 10~--~20\% of those measured by Ardila et al.   The minor residual differences are most likely attributable to the H$_{2}$ removal procedures employed; fitting and removing the contaminating H$_{2}$ lines (this work) versus interpolating over wavelengths contaminated by H$_{2}$ (Ardila et al.). The differences are slightly larger for \ion{N}{5}, where the relative contribution of H$_{2}$ fluorescence is larger.

The inclusion of lower ionization state lines is complicated by abundance variations and absorption from outflows and the interstellar medium (see, e.g., Johns-Krull \& Herczeg 2007).  The strongest magnetospheric lower ionization emission lines in the spectra of WTTSs are \ion{Si}{3} $\lambda$1206 and \ion{C}{2} $\lambda$$\lambda$1334, 1335~\AA.  As noted above, some CTTSs show no silicon emission in their accretion spectra (\ion{Si}{2}, \ion{Si}{3}, \ion{Si}{4}; see also France et al. 2010), possibly due to depletion of refractory elements into grains during the evolution of the disk in the first 10 Myr.\nocite{france10b}  However, other CTTSs show broad \ion{Si}{3} emission lines (e.g., BP Tau), likely formed near the stellar/magnetospheric accretion region.  \ion{C}{2} lines are usually blended with strong H$_{2}$ features and strong ISM+CSM absorption lines, which makes the intrinsic \ion{C}{2} line flux extremely challenging to recover.  In order to assess the relative importance of these low-ions to the total far-UV radiation field strength, we fit the \ion{Si}{3} profiles (when present) and integrated over the 1334~--~1337~\AA\ region covered by \ion{C}{2} because the actual \ion{C}{2} emission lines could not be clearly identified in all cases.  These low-ion line fluxes were then compared to the flux in the \ion{C}{4} doublet.  We find only two stars where $F$(\ion{C}{2})/$F$(\ion{C}{4}) or $F$(\ion{Si}{3})/$F$(\ion{C}{4}) approach unity: HN Tau and RECX-15.  A quick inspection of Table 2 shows that these stars are anomalously weak \ion{C}{4} emitters, therefore the low ions are not particularly strong in these cases, but instead the higher ions are particularly weak.  Excluding these two sources, we find $\langle$$F$(\ion{C}{2})/$F$(\ion{C}{4})$\rangle$~=~0.12~$\pm$~0.07 and $\langle$$F$(\ion{Si}{3})/$F$(\ion{C}{4})$\rangle$~=~0.05~$\pm$~0.04.   Therefore, these emission lines are not included in the compiled radiation fields as they contribute of order 0.2~\% to the total FUV irradiance from the central  stars, although we caution that this number is highly uncertain in the case of \ion{C}{2};  it is possible that this emission line may contribute as much as 1~--~2~\% of the total FUV radiation field.

\subsection{The FUV Continuum Spectrum}

The near-UV continua from CTTS are consistent with shock-generated Balmer continuum emission, and can be modeled to determine the mass accretion rate onto the star when the interstellar reddening is known (e.g. Calvet \& Gullbring 1998; Ingleby et al. 2013).  In TW Hya, the FUV continuum has been shown to be in excess of what is predicted by simply scaling the Balmer continuum to shorter wavelengths~\citep{herczeg04}; contamination by atomic and molecular emission lines and low flux levels make the true FUV continuum essentially impossible to measure without moderate spectral resolution and low instrumental backgrounds for all but the brightest CTTSs.  Previous studies of the FUV continuum have suggested two primary origins for this emission; a hot accretion component~\citep{costa00,herczeg04,france11a} and an H$_{2}$ molecular dissociation quasi-continuum generated by collisions with non-thermal electrons~\citep{bergin04,herczeg04,ingleby09}.   \citet{france11a,france11b} presented the first detailed observations of the FUV continuum, concluding that while a combination of accretion continuum, electron-impact H$_{2}$ emission, and CO fluorescence (excited by Ly$\alpha$ and \ion{C}{4} photons) can reproduce some features of the spectra, there are significant discrepancies between the predicted and observed spectral features from this process.  

\begin{figure}
\figurenum{A.4}
\begin{center}
\epsfig{figure=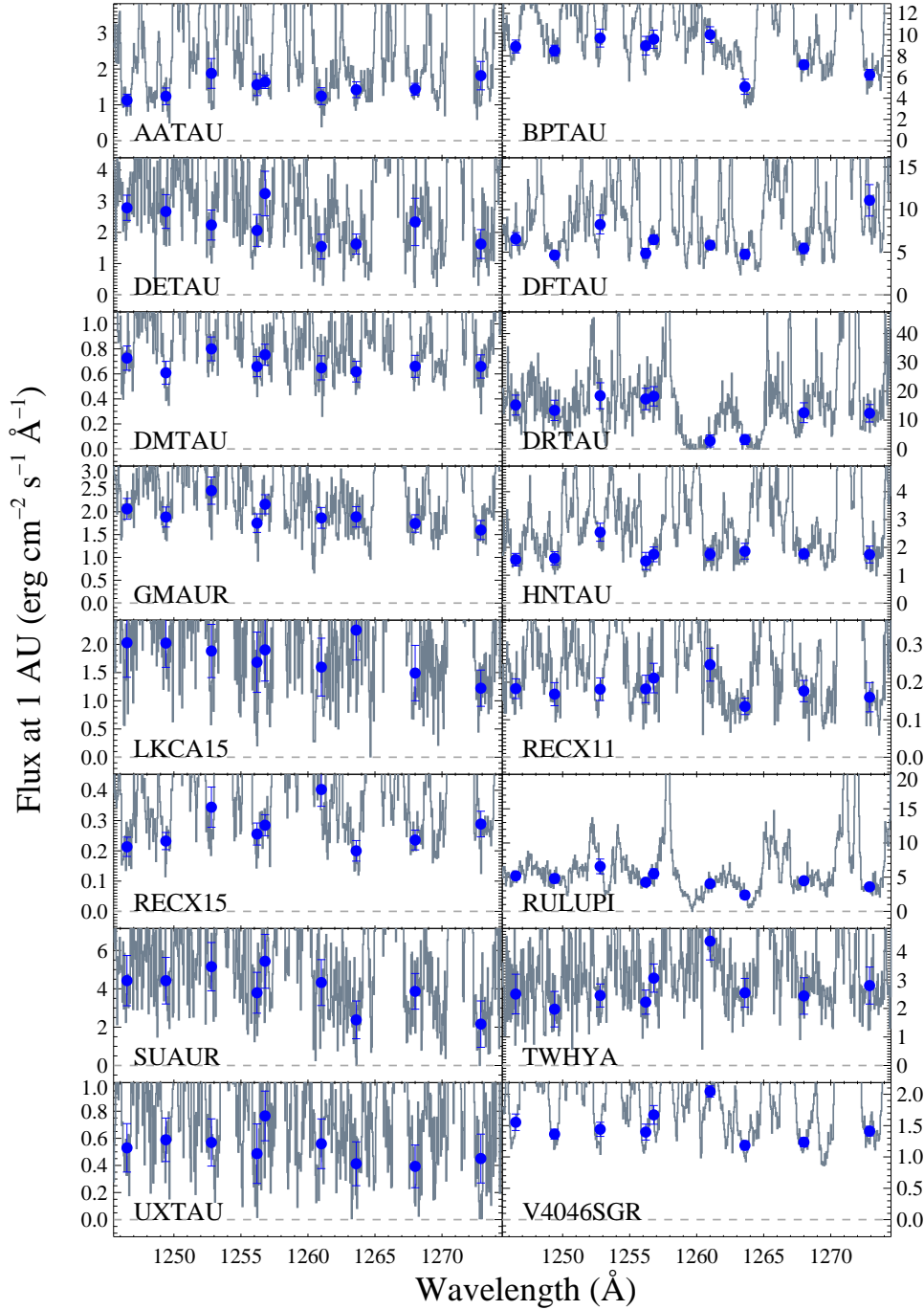,width=4.5in,angle=00}
\vspace{+0.35in}
\caption{A detailed look at how the binned FUV continuum points are defined.  Each bin is the average flux over a 0.75~\AA\ emission line-free region of the spectrum (blue filled circles) and the uncertainty is defined as the standard deviation about that mean flux.  Absorption from low-ionization atomic species (in this case \ion{Si}{2} $\lambda$$\lambda$1260,1264) is observed in the spectra of DR Tau and RU Lupi (see also Herczeg et al. 2005), most likely from low-ionization outflows in the circumstellar environment.  
\label{cosovly}}
\end{center}
\end{figure}

\begin{deluxetable}{l|cc|cc}
\tabletypesize{\footnotesize}
\tablecaption{Table A.1~--~Average FUV Conintuum Flux at 1 AU and Signal-to-Noise per Bin\label{lya_targets}}
\tablewidth{0pt}
\tablehead{ \colhead{Target}   &   \colhead{$\langle$$F\tablenotemark{a}$(1140~--~1340\AA)$\rangle$}    &   \colhead{$\langle$$S/N\tablenotemark{b}$(1140~--~1340\AA)$\rangle$}    &     \colhead{$\langle$$F\tablenotemark{a}$(1660~--~1700\AA)$\rangle$}      &   \colhead{$\langle$$S/N\tablenotemark{b}$(1660~--~1700\AA)$\rangle$}    \\ 
   & (erg cm$^{-2}$ s$^{-1}$ \AA$^{-1}$)  &        &           (erg cm$^{-2}$ s$^{-1}$ \AA$^{-1}$)        &     } 
\startdata
AA Tau    &    1.2   &  5.0    & 2.1  &    4.8     \\  
BP Tau    &    9.2  &  11.5    &  14.8  &  13.9          \\  
DE Tau    &   2.6   &   3.7    &  2.2  &    2.8        \\  
DF Tau    &    6.0   &   7.4     &  13.6  &   10.3       \\  
DM Tau    &    0.7   &  6.3     &  0.8  &   4.9       \\  
DR Tau    &    19.5  &  3.8     &  42.1  &    6.4       \\  
GM Aur    &    2.1  &   7.8     &  2.7  &  5.6          \\  

HN Tau    &    1.9   &  5.9   &  3.4      & 7.0      \\  

LkCa 15    &    2.0  &  3.1    &  1.9  &    2.7      \\  
RECX 11    &    0.2   &   4.8     &  0.2  &    4.1      \\  
RECX 15    &    0.2   &   5.5    &  0.2  &    4.7    \\  
RU Lupi    &    5.2   &  7.1    &  18.9  &    17.0  \\  

SU Aur    &    4.9   &   2.9     &  5.2  &    3.0          \\ 
TW Hya    &   2.2   &   3.3     &  4.9  &    4.4  \\ 

UX Tau    &    0.5   &   2.5     &  0.4  &    0.9   \\  
V4046 Sgr    &    1.6  &   14.8    &  1.1  &   12.4  
\enddata
\tablenotetext{a}{Average reddening corrected flux in the measured, binned continuum spectra (\S A.3) over the indicated bandpasses.  Fluxes are evaluated at 1 AU from the target star for comparison with Figures A.1 and A.3~--~A.6. }
\tablenotetext{b}{Average signal-to-noise ratio per binned continuum point over the indicated bandpasses.  }  

\end{deluxetable}

We have created new FUV continuum spectra for the 16 targets considered in this study.   Owing to the various H$_{2}$ and CO pumping transitions populated in different targets, each CTTS molecular emission spectrum is essentially unique.  In order to find true continuum regions, we searched the spectra by hand and identified 210 spectral points where a 0.75~\AA\ (approximately 10 spectral resolution elements) emission line-free window can be identified between 1138 and 1791~\AA.  We refer to these as `binned FUV continuum spectra'.  Each individual binned FUV continuum point is taken as the mean of the flux in this $\pm$~0.375~\AA\ spectral window free of molecular and atomic line emission.  The error on each individual binned FUV continuum point is taken to be the standard deviation about the mean flux in the $\pm$~0.375~\AA\ spectral window.  A comparison of the observed spectra and the binned FUV continua spectra is shown in detail for the 1245~--~1275~\AA\ spectral region in Figure A.4.  The binned continuum spectra are then corrected for interstellar reddening, assuming the optical extinctions given in Table 1 and $R_{V}$~=~3.1.  

 In order to estimate the true underlying continuum, we fitted the spectra in three broad regions free from residual features in the binned FUV continuum spectra (over the intervals $\Delta$$\lambda$~=~1160~--~1190~\AA, 1245~--~1330~\AA, and 1663~--~1685~\AA) with a second order polynomial.  In Figure A.5, we show the short-wavelength portion of the binned FUV continuum spectrum with the polynomial fit overplotted as the red dashed line.    In Table A.1, we present the average flux and the average S/N ratio per binned continuum point for each target in the sample, in both short-wavelength (1140~--~1340~\AA) and long-wavelength (1660~--~1700~\AA) bands free of molecular continuum emission.   While the average S/N ratio per binned continuum point is $>$~2.5 for all stars in the short-wavelength portion of the $HST$ bandpass used in this work (1140~--~1340~\AA), we caution that the data quality of the continuum for the fainter targets (DE Tau, DR Tau, LkCa15, SU Aur, TW Hya, and UX Tau A) makes the short-wavelength continuum fitting less reliable.  In these cases, simply taking the average FUV continuum level may be more appropriate than the polynomial fit to the FUV continuum.   

\begin{figure}
\figurenum{A.5}
\begin{center}
\epsfig{figure=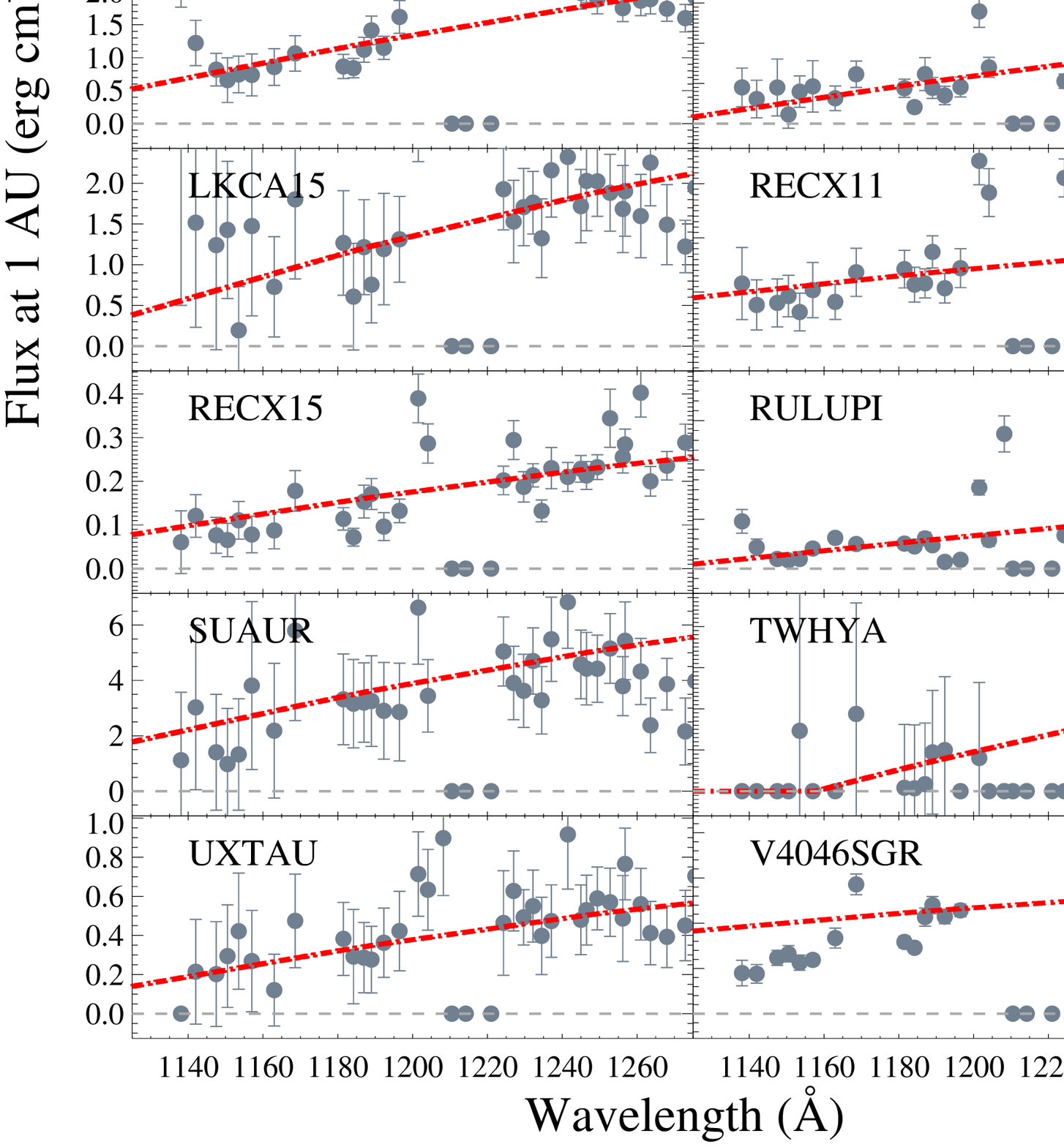,width=4.5in,angle=00}
\vspace{+0.35in}
\caption{The binned FUV continuum spectra at the shortest wavelength in our $HST$ data set (1140~--~1275~\AA).
The center of the Ly$\alpha$ emission line is contaminated by geocoronal emission in most cases and has been removed, resulting in the zero-flux points from 1210~--~1220~\AA.  The binned FUV continuum points have average S/N per point $>$ 2.5 for all targets in this range and the polynomial fit to the FUV continuum is shown as the red dash-dot line. TW Hya displays several points as zeroes in this wavelength range because of the truncated bandpass and lower sensitivity of STIS (relative to COS) at $\lambda$~$<$~1250~\AA.  
\label{cosovly}}
\end{center}
\end{figure}

\begin{figure}
\figurenum{A.6}
\begin{center}
\epsfig{figure=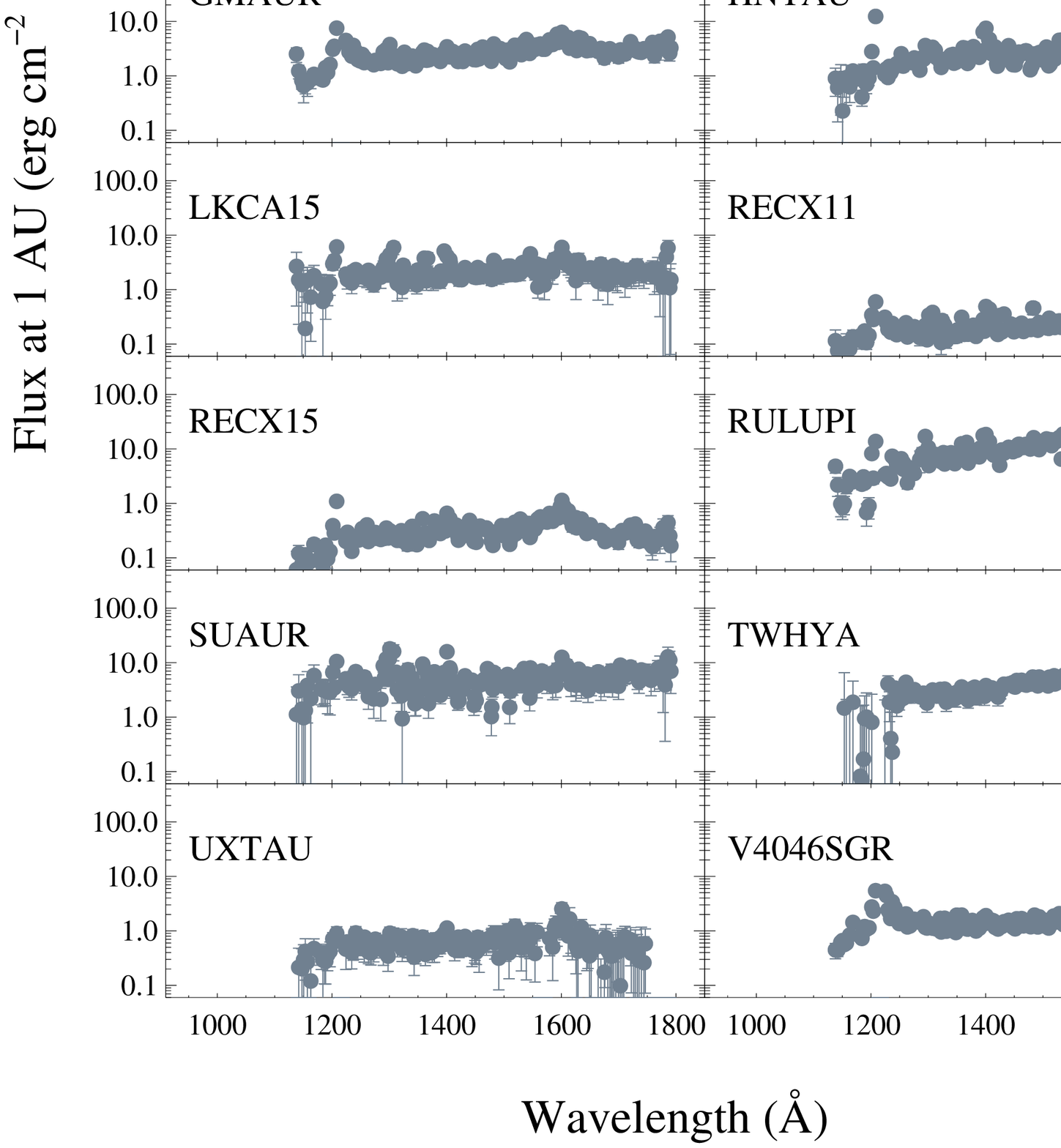,width=4.5in,angle=00}
\vspace{+0.5in}
\caption{The binned FUV continuum spectra, shown on a common logarithmic scaling.  Each bin is the average flux over a 0.75~\AA\ emission line free region of the spectrum.  Objects with a significant number of points below the continuum level indicate absorption from a CO-rich circumstellar disk~\citep{france12a,mcjunkin13} or atomic outflow~\citep{krull07}. 
\label{cosovly}}
\end{center}
\end{figure}

We have previously described the use of a first order linear fit~\citep{france11a}, but we employ a curvature term here to account for uncertainty in the reddening correction.  While most of our spectra maintain appreciable signal-to-noise to $\lambda$~$\sim$~1760~\AA, we truncate the fits at 1685~\AA\ to accommodate the long wavelength cut off of the STIS E140M data used for TW Hya.  The binned continua are shown on a linear scale, with the polynomial continuum fits overplotted, in Figure 2 and the extrapolation of our continuum fit to $\lambda$~$<$~1140~\AA\ is shown compared with archival $FUSE$ observations in Figure~3. \nocite{mcjunkin13} 

The complete binned FUV continuum spectra are shown scaled to the flux at 1 AU in Figure A.6.  There are several residual features in this continuum spectrum: $(i)$ emission excess between 1200~--~1230~\AA\ that may be a very broad ($\pm$~4000 km s$^{-1}$) Ly$\alpha$ component, $(ii)$ CO $A$~--~$X$ absorption bands  between 1400~--~1520~\AA\ (e.g., AA Tau and DE Tau; McJunkin et al. 2013), 
$(iii)$ absorption from low-ionization atomic species most likely arising in outflows (e.g. DR Tau and RU Lupi; Herczeg et al. 2005),\nocite{herczeg05} and $(iv)$ an excess emission feature spanning 1520~--~1660~\AA\ in some targets (e.g., AA Tau, DM Tau, GM Aur, RECX-15, V4046 Sgr).  This latter feature is the ``1600~\AA\ Bump'' and we will present a detailed analysis of this quasi-continuous emission feature in a future work.  Figure A.5 shows the binned spectra at the short-wavelength end of the $HST$ bandpass; broad Ly$\alpha$ emission lines can be seen in the binned continuum spectra.  The center of the Ly$\alpha$ emission line is contaminated by geocoronal emission in most cases and has been removed, resulting in the zero-flux points seen in from 1210~--~1220~\AA\ in Figure A.5.


\bibliography{ms}

\end{document}